%% file: FH_Manuscript.tex
\documentclass[twocolumn]{aastex63}

\usepackage{url}

\turnoffedit

\accepted{October 26, 2022}

\shorttitle{Structure and History of Fornax-Horologium}
\shortauthors{Kerr et al.}

\begin{document}

\title{SPYGLASS. III. The Fornax-Horologium Association and its Traceback History within the Austral Complex}

\correspondingauthor{Ronan Kerr}
\email{rmpkerr@utexas.edu}

\author[0000-0002-6549-9792]{Ronan Kerr}
\affiliation{Department of Astronomy, University of Texas at Austin\\
2515 Speedway, Stop C1400\\
Austin, Texas, USA 78712-1205\\}

\author[0000-0001-9811-568X]{Adam L. Kraus}
\affiliation{Department of Astronomy, University of Texas at Austin\\
2515 Speedway, Stop C1400\\
Austin, Texas, USA 78712-1205\\}

\author[0000-0002-5648-3107]{Simon J. Murphy}
\affiliation{Centre for Astrophysics, University of Southern Queensland\\
Toowoomba, QLD 4350 Australia.\\}

\author[0000-0001-9626-0613]{Daniel M. Krolikowski}
\affiliation{Department of Astronomy, University of Texas at Austin\\
2515 Speedway, Stop C1400\\
Austin, Texas, USA 78712-1205\\}

\author[0000-0001-5222-4661]{Timothy R. Bedding}
\affiliation{Sydney Institute for Astronomy, School of Physics, University of Sydney NSW 2006, Australia}

\author{Aaron C. Rizzuto}

\begin{abstract}

The study of young associations is essential for building a complete record of local star formation processes. The Fornax-Horologium association (FH), including the $\chi^1$ Fornacis cluster, represents one of the nearest young stellar populations to the Sun. This association has recently been linked to the Tuc-Hor, Carina, and Columba associations, building an extensive ``Austral Complex'' almost entirely within 150 pc. \edit1{Using Gaia astrometry and photometry in addition to new spectroscopic observations, we perform the deepest survey of FH to date, identifying over 300 candidate members, nearly doubling the known population.} By combining this sample with literature surveys of the other constituent populations, we produce a contiguous stellar population covering the entire Austral Complex, allowing the definitions of sub-populations to be re-assessed along with connections to external populations. This analysis recovers new definitions for FH, Tuc-Hor, Columba, and Carina, while also revealing a connection between the Austral complex and the Sco-Cen-affiliated Platais 8 cluster. This suggests that the Austral complex may be just a small component of a much larger and more diverse star formation event. Computing ages and tracing stellar populations back to formation reveals two distinct nodes of cospatial and continuous formation in the Austral Complex, one containing Tuc-Hor, and the other containing FH, Carina, and Columba. This mirrors recent work showing similar structure elsewhere, suggesting that these nodes, which only emerge through the use of traceback, may represent the clearest discrete unit of local star formation, and a key building block needed to reconstruct larger star-forming events.  

\end{abstract}


\section{Introduction} \label{sec:intro}

Young clusters and associations provide critical information on how star formation progresses over large timescales \citep{Krumholz19, Adamo20}. While active sites of star formation can provide remarkable insight into the dynamics at play during these processes, they provide only snapshots, barring the study of processes occurring continuously throughout the star formation process, and making research on rare or brief star-forming events much more challenging. The study of young associations is therefore highly complementary, as these groups of recently-formed stars hold a record of star formation that spans tens of millions of years before present. Through the dynamics of stars in these associations, the nature of the cloud that birthed them and the processes that guided their formation can be reconstructed, providing a result that can truly track the entire history of a star forming event, from the formation of the first dense cloud to when the gas finally disperses. Since these studies generally require accurate velocity data for each candidate member, large-scale studies of associations have long been impractical, however Gaia astrometry and high-precision ground-based RVs are making these critical stellar properties more accessible than ever before \citep[e.g.,][]{Wright22}.

The Gaia data releases \citep{GaiaMission, GaiaDR218, EDR3Astro_Lindegren21, GaiaDR3_22} have recently driven discoveries in star formation and stellar populations. By providing essentially complete 3-D positions and proper motions among stars with 12$<$G$<$17, overdensities in not just spatial coordinates, but also velocity coordinates can be identified \citep{EDR3Astro_Lindegren21}. This data set has already been used for the large-scale detection of star clusters and moving groups extending out to over 1 kpc away from the Sun \citep[e.g.,][]{Kounkel19, Sim19,Prisinzano22}. Gaia photometry can also been used to identify young stars that have not yet fully contracted onto the main sequence, allowing for young populations to be isolated from a complex field and permitting studies targeting young populations without the severe field contamination that would otherwise decimate the sensitivity of any clustering efforts \citep[e.g.,][]{Zari18, CantatGaudin19b}. The SPYGLASS program (Stars with Photometrically Young Gaia Luminosities Around the Solar System) takes advantage of these unique strengths from Gaia, isolating young stars through a Bayesian identification framework, and performing sensitive spatial and kinematic clustering on the photometrically young stars that are identified. \citet{Kerr21} (hereafter referred to as SPYGLASS-I) outlined the process for detecting young stars while also publishing a catalog of candidate young stars within 333 pc. That work identified 27 young clusters and associations, including some features that have been overlooked in other similar works. 

The group identified as ``Fornax-Horologium'' in SPYGLASS-I has the nearest average distance ($d = 108$ pc) of any top-level group identified in that work, representing an extremely accessible population for further study. Furthermore, it has a unique position within the current literature. Initially identified through its dense core under the name Alessi 13 \citep{Dias02} and later referred to as the ${\chi}^1$ Fornacis Cluster \citep{Mamajek16}, the group received little coverage between then and recent times, mainly being featured as part of large catalogs and bulk investigations covering many other nearby groups. The works since discovery include age estimates for the group \citep{Kharchenko05, Yen18}, alongside the refinement of basic cluster parameters and expansion of the known population to 48 members \citep{CantatGaudin18}. However, with \citet{Kharchenko05} and \citet{Yen18} both computing ages in excess of 500 Myr, there was little interest in this cluster among those interested in young stars during this time period. A younger age for the group has only recently emerged, with \citet{Mamajek16} noting a presence of X-ray excesses that suggest an age just over 30 Myr. Recent works now strongly support this younger age range for the ${\chi}^1$ Fornacis cluster, with recent isochronal estimates from SPYGLASS-I, \citet{Zuckerman19}, and \citet{Galli21} all supporting an age between 30 and 40 Myr. 

Some nearby young associations have recently been compared to this group due to similarities in not just age, but also motion, causing \citet{Zuckerman19} to suggest a common origin with the Tuc-Hor and Columba associations, and \citet{Mamajek16} and \citet{Galli21} to suggest the additional inclusion of the Carina association in that same common formation complex. A similar conclusion had previously been reached by \citet{Torres01}, causing the complex, excluding $\chi^1$ Fornacis, to be collectively dubbed the ``Great Austral Young Association'' (GAYA), which we abbreviate to the ``Austral Complex'' to avoid confusion with ``Gaia''. The emergence of well-defined kinematic differences between the three regions resulted in this unified view becoming less popular \citep[e.g.,][]{Torres08}. However, \citet{Galli21} recently showed that, while these associations do have subtly different velocities, they shared a much closer configuration around the time of formation, suggesting that the existence of these groups may be explained through common formation and dispersal from one another. 

\begin{figure}
\centering
\includegraphics[width=7.8cm]{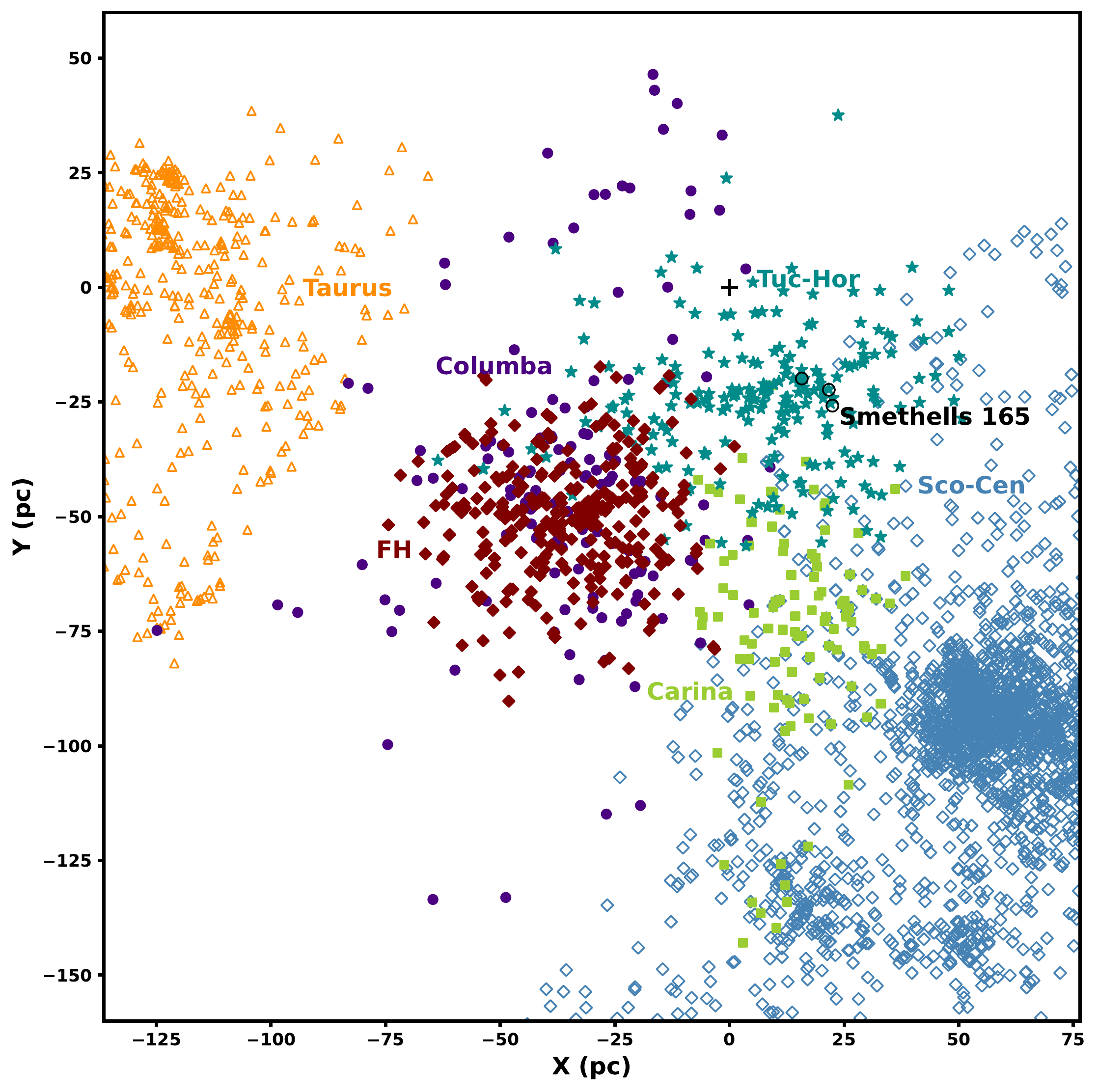}\hfill
\caption{The layout of the Austral complex, as we later construct in Sections \ref{sec:data} and \ref{sec:memselect}. FH members are based on our own novel sample (see Sec. \ref{sec:gaiacansel}), while Carina, Columba, and Tuc-Hor members are drawn from literature (see Sec. \ref{sec:thcc}). We also include the notable surrounding associations of Sco-Cen and Taurus as shown in SPYGLASS-I, which may have connections to the Austral Complex, as well as proposed members of the Smethells 165 Association. All groups are labelled, and the sun is included as a black cross for reference. Columba, Carina, Tuc-Hor, and FH all form a largely contiguous population while parts of Sco-Cen overlap with these populations, especially Carina.}
\label{fig:FHsurroundings}
\end{figure}

New populations and connections to the Austral complex have been recently proposed, most notably \citet{LeeSong19} linking 32 Orionis to Columba, the former being previously identified as part of Taurus in SPYGLASS-I, and Smethells 165 \citep{Higashio22}, which is a proposed moving group defined using three co-moving stars otherwise closely connected to Tuc-Hor. \edit1{\citet{Gagne21} also recently proposed links between Tuc-Hor and IC 2602 and between Platais 8, Carina, and Columba, which connects these Austral populations to sub-clusters within the SPYGLASS-I extent of the Sco-Cen association. We provide a visual representation of the neighborhood surrounding FH in Figure \ref{fig:FHsurroundings}, showing the contiguous distribution formed by FH, Tuc-Hor, Columba, and Carina, and the proximity of potentially connected components of Taurus and Sco-Cen.} However, the possibility of connected star formation between these groups and $\chi^1$ Fornacis has yet to be full explored. 

There has yet to be an attempt to firmly establish a unified formation explanation that covers this entire star formation complex. Such a new view of star formation in the region can re-contextualize proposed links in terms of plausibility of a common formation origin, rather than just common velocity. Recent work through SPYGLASS-I, as well as a targeted study of the Cepheus Far North association (CFN) in \citet{Kerr22} (hereafter SPYGLASS-II), have demonstrated the existence of star-formation patterns that include multi-origin star forming events spanning over 10 Myr, including CFN and Perseus OB2. With three established populations potentially connected to ${\chi}^1$ Fornacis, all of which having extensive past spectroscopic coverage, this cluster and its surroundings provides a unique opportunity to investigate a nearby and accessible population with strong age constraints. Our sensitive membership detection, enabled through the methods displayed in SPYGLASS-I, also enable the further expansion of the cluster's known population into its more tenuous outskirts, surpassing the 108 members identified in \citet{Zuckerman19} and 164 members in \citet{Galli21} and thereby producing a rich sample through which further studies can be conducted.

Using a Gaia DR3-updated version of the Fornax-Horologium association, as presented in SPYGLASS-I, we perform the most detailed analysis of the population to date, combined with a broader analysis including the entire Austral complex. We introduce the data used in this paper in Section \ref{sec:data}, \edit1{which include a deepened Gaia survey combined with 37 new spectroscopic observations from the Las Cumbres Observatory (LCO), allowing for the careful assessment of the youth, membership, and dynamics of candidate members using radial velocities and lithium EWs. We assess} the membership of both new FH candidate members and literature Austral members in Section \ref{sec:memselect}. We then perform our main analysis in Section \ref{sec:results}, re-clustering on the combined Austral population, computing ages, and performing traceback on the resulting populations. Section \ref{sec:stellarpops} provides an expanded stellar census for FH. We then discuss the implications of the patterns seen across the Austral complex in Section \ref{sec:discussion}, before concluding in Section \ref{sec:conclusion}.

\section{Data} \label{sec:data}

\subsection{Gaia Candidate Selection} \label{sec:gaiacansel}

The initial selection of candidate Fornax-Horologium members is based on the results of SPYGLASS-I, which identified 3$\times$10$^4$ photometrically young stars within 333 pc of the solar system. SPYGLASS-I used the HDBSCAN clustering algorithm to identify young populations out of this sample, and grouped 40 of these proposed young stars together as the Fornax-Horologium association, which represents the closest top-level group to the solar system identified in that work. However, with a SPYGLASS-I age at the upper end of that paper's $\la$ 50 Myr sensitivity range, we expect a relatively low recovery rate of members of between 10 and 20\%. As such, a more permissive sample of possible members is required to produce a reasonably complete view of the association.

Our methods for producing this more complete view were also employed in SPYGLASS-II. We re-ran the SPYGLASS-I analysis using the updated Gaia DR3 sample which, in the context of our analysis, is the same as the EDR3 sample used in SPYGLASS-II but with the addition of new radial velocity data. Using this sample, we identified stars with photometry indicative of youth, and then applied the HDBSCAN clustering algorithm to that young sample, extracting the clump associated with the Fornax-Horologium association. We then searched for phase space neighbors, which reintroduced many genuine members that were missed due to them having already evolved to the main sequence.

This process increased the number of photometrically young stars grouped into Fornax-Horologium to 48, which were accompanied by 654 space-velocity neighbors. Our identification of neighbors used looser quality requirements than the sample for cluster identification. We only removed targets without a Gaia $G$ magnitude, which is necessary to assess a target's feasibility for follow-up observations, or without a 5-parameter astrometric solution from Gaia, which ensures that all sources have well-constrained motions in at least two dimensions to properly assess whether they are co-moving with known Fornax-Horologium members. While we did not introduce any additional restrictions on Gaia photometric or astrometric quality for our broader sample restriction, we did include flags for the quality of the Gaia astrometric and photometric solution, which match the flags used in SPYGLASS-II. 

The population produced by our selection was relatively uncontaminated, so no additional cuts were needed to achieve a more equitable balance of members and field stars, unlike for CFN in SPYGLASS-II, where a restriction of the clustering proximity parameter ($D$; ``strength'' in SPYGLASS-I) was required to avoid a field-dominated candidates sample. A more detailed description of the young member identification and nearby candidate selection process is provided in SPYGLASS-II. As we explore further in Section \ref{sec:fhpops}, the population we have produced is considerably broader than previous coverage of the region when it was referred to as Alessi 13 and ${\chi}^1$ For, and includes sections of the association that, unlike the central cluster, are not plausibly bound. This motivates the continued use of the name Fornax-Horologium (FH) to refer to the entire population including the unbound extended halo, a convention we maintain throughout this paper. Meanwhile, ${\chi}^1$ For remains our preferred manner of referring to the central clustered core around the namesake stellar system. 

\subsection{Literature Radial Velocities}

Radial velocities are a critical component in establishing membership of stars within young populations, especially above the pre-main sequence turn-on at $G \lesssim 8$, where field stars are photometrically indistinguishable from members. Despite Gaia DR3 RVs covering most stars with apparent magnitudes $G < 14.5$ \citep{GaiaDR3_RVs_Katz22}, including 277 members of our Fornax-Horologium candidate list, only 91 of that list's 654 objects have sub-km s$^{-1}$ uncertainty. Furthermore, RV measurements taken as part of SPYGLASS-II often showed slight inconsistencies with Gaia radial velocities at the km s$^{-1}$ level, despite smaller uncertainties, producing anomalous traceback patterns when using Gaia compared to the very consistent results attained through independent radial velocity measurements. The incomplete nature of the Gaia sample and apparent inconsistency of Gaia RVs highlight the need for higher-quality radial velocity measurements across the association. While most of these Gaia RVs are sufficient for assessing the membership of a star, reliable RV measurements with sub-km s$^{-1}$ precision are necessary for accurate traceback and other forms of kinematic analysis

To improve our radial velocity coverage, we collected additional measurements from Simbad and Vizier to improve the completeness of the radial velocity sample, keeping the lowest-uncertainty measurement in the literature. If a measurement aside from Gaia had $\sigma_{RV} < 1$ km s$^{-1}$, it was always selected over the Gaia results due to those aforementioned inconsistencies. This process located new RVs for 26 stars. Our final literature RV sample therefore consisted of 303 stars, 277 from Gaia DR3 \citep{GaiaDR3_22}, and the remaining 26 from external sources \citep{Gontcharov06, Kharchenko07, Casagrande11, deBruijne12, Moor13, Elliot14, RAVEKunder17, Shkolnik17, Gagne18BXIII, Schneider19, Steinmetz20}. A total of 106 stars in this sample have uncertainties of $<1~$km s$^{-1}$ which, while useful, still leaves 32 stars with $G<13$ without such measurements. 

\subsection{New Spectroscopic Observations from LCO}

Fornax-Horologium's incomplete literature RV sample and previously-noted internal inconsistencies in the Gaia measurements, which represent the majority of that sample, highlight the need for new observations to enable dynamical studies. Such observations are also useful for measuring spectral line indicators such as Li 6708\AA~absorption and H$\alpha$ emission, which can both be used to firmly establish the youth of a star. We obtained these supplemental observations using the Las Cumbres Observatory (LCO) NRES spectrograph, which provides high-resolution  ($R=53000$) spectra covering a spectral range between 3400 and 10900 \AA. Exposure times ranged from 5 to 30 minutes depending on the star's magnitude. These exposure times were typically chosen to ensure $S/N=30$  around the Li 6708 \AA\ line spectral window, which permits for robust measurement of that line. We allowed the S/N ratio in that spectral window to drop as low as 10 for stars near the $G\sim13$ sensitivity limit of the telescope, where reliable Li measurements are generally not attainable but sub-km s$^{-1}$ RVs are still possible.

Observations covered every candidate without a pre-Gaia DR3 radial velocity with a magnitude $G < 13$, as well as a subset of brighter candidate members with larger uncertainties in literature. Three stars with $G < 13$ had a good Gaia DR2 RV, which resulted in the choice not to observe them, but did not have an RV solution in Gaia DR3. These stars are the only stars on the main sequence without RV coverage after our observations, as we select Gaia DR3 solutions over DR2 even in the case of a null solution. A total of 37 targets were observed using the LCO 1m telescopes across three of the network's nodes: Wise Observatory (TLV), South Africa Astronomical Observatory (CPT), and Cerro Tololo Observatory (LSC). NRES data products are automatically reduced through the BANZAI data reduction pipeline \citep{McCullyBANZAI18}, and the resulting data products were used directly for further analysis.

\subsection{RVs, Lithium and H\texorpdfstring{$\alpha$}{Lg}}

We computed radial velocity measurements using the spectral line broadening functions from the {\tt saphires} packages \citep{Tofflemire19}, yielding radial velocity measurements with sub-km s$^{-1}$ uncertainty for 23 of the stars. Most stars with poorer RV uncertainties had highly broadened line profiles and, as such, better constraining these measurements would likely require significantly more signal than was accessible at LCO. One star received duplicate observations and, in that case, we recorded the weighted average of those two measurements. Final RV measurements for all candidate members are provided in Table \ref{tab:spectres}. 

Our NRES spectra can also be used to assess the youth of Fornax-Horologium members using spectral indicators, most notably Li 6708\AA~absorption, which depletes early in the life of stars, and H$\alpha$, which is typically associated with accretion and is therefore especially common in later-type young stars. Both lines are covered in their entirety by NRES, so equivalent width (EW) measurements were attempted for all objects observed. Our EW measurements used a simple least-squares optimization routine on a Gaussian line profile, normalized such that the background is at 1. We did not deblend our Li EWs with the nearby 6707.4 \AA~Fe I lines. This may produce measurement discrepancies on the scale of 10--20 m\AA, however the effects of this should typically be smaller than our EW uncertainties. The resulting values are presented in Table \ref{tab:spectres}. Some example fits to both RV and EWs are provided in Figures 1 and 2 of SPYGLASS-II. 

\subsection{Related Associations} \label{sec:thcc}

Due to our addition of numerous new observations covering the Fornax-Horologium Association, we have the ability to assess its motions more clearly than before. Those motions enable a detailed study of its position relative to other nearby and potentially related associations, especially Tucana-Horologium, Columba, and Carina, with which FH has been previously connected \citep[e.g.,][]{Torres01, Mamajek16, Galli21}. Other associations, most notably 32 Orionis, have also been suggested as connected \citep[e.g.,][]{LeeSong19} but, as we show in Section \ref{sec:exttraceback}, these other associations do not show orbits and ages consistent with related formation. While the projection effects on transverse velocity caused by these associations' proximity to the Sun makes our general SPYGLASS framework unsuitable for performing our own independent membership census covering them, literature coverage of these associations already provides extensive membership lists, as well as radial velocities and spectral youth indicators for a significant subset of their members. By combining these literature sources, we can produce a sample with completeness suitable for analysis alongside our coverage of FH using SPYGLASS methods. 

The earliest catalog with significant coverage of all three of these associations was produced by \citet{Torres08}, which identifies 44 Tuc-Hor members, 41 Columba members, and 23 Carina members. There are no RVs reported in this paper, but it did make use of previous catalogs such as \citet{Torres06} to provide RVs for membership assessments, which can be reintroduced later through a search of the literature. The sample of members of these associations was later expanded in \citet{Malo13}, which provides lists of both \textit{bona fide} members and candidates for all three associations derived using a Bayesian identification tool. The paper's catalog includes 44 members and 26 candidates for Tuc-Hor, 21 members and 26 candidates in Columba, and 5 members and 6 candidates in Carina. All \textit{bona fide} members have RV coverage, drawn from a variety of sources, while about half of candidate members do, a subset of which also have lithium measurements. Our sample for Tuc-Hor can be significantly deepened using \citet{Kraus14}, which provides a significant expansion to past observational work in the association, spectroscopically confirming 129 members of Tuc-Hor and providing additional RV and Li measurements for 13 previously known members. Finally, we added objects identified by \citet{Schneider19}, which used new spectroscopic observations to confirm the membership of 77 candidate members of many nearby associations, including the three that interest us. This provides an additional 7 objects with full RV and spectral coverage in Tuc-Hor, 11 in Columba, and 10 in Carina.  %Samples from Argus were also available through \citet{Malo13} and \citet{Schneider19}, which have similar proposed ages to the other associations we discuss here. However, we excluded them from further analysis upon determining that their motions are not consistent with ${\chi}^1$ Fornacis. 

These sources all provide a rich collection of spectroscopic data, although they still lack completeness among fainter stars, especially in Carina and Columba. To improve our sample's completeness for non-kinematic applications, we included the catalogs from \citet{Gagne18BXI},  \citet{Gagne18BXII}, \citet{Gagne18BXIII}, which together provide hundreds of additional likely members for all three associations we consider, including RVs from literature sources. These stars were identified using the BANYAN $\Sigma$ Bayesian classification algorithm, which identifies prospective association members using a spatial and kinematic model of the association. We also included an additional sample from \citet{Booth21}, which used the BANYAN algorithm to expand the sample of probable members in Carina, adding four candidates not identified elsewhere. 

We combined all of these literature samples for Tuc-Hor, Columba, and Carina into a single master list, and added photometric and astrometric data from the Gaia DR3 catalog. We removed duplicates, along with any stars that do not have the 5-parameter Gaia astrometric solution necessary to analyse their motions. This merged sample contains 514 stars, including 237 in Tuc-Hor, 159 in Columba, and 118 in Carina. Finally, we performed a search of Simbad and Vizier for RV sources not included so far, recovering new RVs from a wide assortment of sources \citep{Grenier99,Bobylev06,Gontcharov06,Torres06,White07,Kharchenko07, Casagrande11,deBruijne12,Anderson12,LopezMarti13,Moor13,Elliot14,Malo14,Burgasser15,Desidera15,Faherty16,RAVEKunder17,Shkolnik17, Fouque18,Gagne18BXI,Gagne18BXII,Gagne18BXIII,Jeffers18,Kounkel18,Soubiran18,Flaherty19,Sperauskas19,Xiang19,Steinmetz20}. This search provides excellent coverage in these FH-associated regions, with 431 of 514 sources having radial velocities. We also collected additional Lithium and H$\alpha$ data from outside our base sample. The sources included are \citet{Riaz06} and \citet{Gizis02}, which provide H$\alpha$ line measurements, \citet{Mentuch08} and \citet{daSilva09}, which provide Lithium line measurements, and \cite{Torres06}, \citet{Kiss11}, \citet{Rodriguez13}, \citet{Bowler19}, which provide both spectroscopic youth indicators. The addition of these sources results in spectral line coverage for nearly half of our sample. 

While there are likely some members falsely assigned to a group, especially for sources identified using the BANYAN algorithm which do not generally have spectroscopic coverage, we find that all associations provide relatively contamination-free samples on the color-magnitude diagram. This suggests that interlopers can be largely ignored, although culling of extreme photometric outliers can still be useful. The depth and apparent purity of this sample makes it an internally consistent addition to our SPYGLASS sample for Fornax-Horologium, which we later combine to form an aggregate catalog for the entire Austral Complex.

\subsection{TESS photometry}

\edit1{To determine pulsation frequencies for asteroseismic ages, we used photometry from the Transiting Exoplanet Survey Satellite (TESS; \citealt{rickeretal2015}). The data have 2-min cadence and were obtained from the Mikulski Archive for Space Telescopes (MAST)\footnote{The specific observations analyzed can be accessed via \dataset[10.17909/t9-nmc8-f686]{https://doi.org/10.17909/t9-nmc8-f686}} using the {\sc lightkurve} package 
\citep{lightkurvecollaboration2018}. No post-processing was applied to these light curves. The identification of stellar pulsation modes in these data and the calculation of stellar pulsation models are described in Sec.\,\ref{sssec:seismology}.}

\subsection{Stellar Masses}

Stellar masses are important for assessing both the past and current environment in which Fornax-Horologium resides. A sufficiently high total stellar mass has the potential bind the association, making it an open cluster in which gravity overcomes the internal motions of the association. Gravitational binding may also be able to deflect the orbits of stars passing through the association, possibly disrupting the traceback of stars that pass though it. To assess the masses in Fornax-Horologium, we used the same isomass tracks for computing stellar masses that we used in SPYGLASS-II, which were derived from PARSEC v1.2S isochrones. We regridded those isochrones onto a grid with mass intervals at every 0.005 M$_\odot$ between 0.09 and 1 M$_\odot$, every 0.01 M$_\odot$ between 1 and 2 M$_\odot$, and every 0.02 M$_\odot$ between 2 and 4 M$_\odot$, and assigned the mass of the nearest model track to each star.

\input{AUSTRAL_SPECTRA}

\section{Member Selection} \label{sec:memselect}

\begin{figure}
\centering
\includegraphics[width=8.0cm]{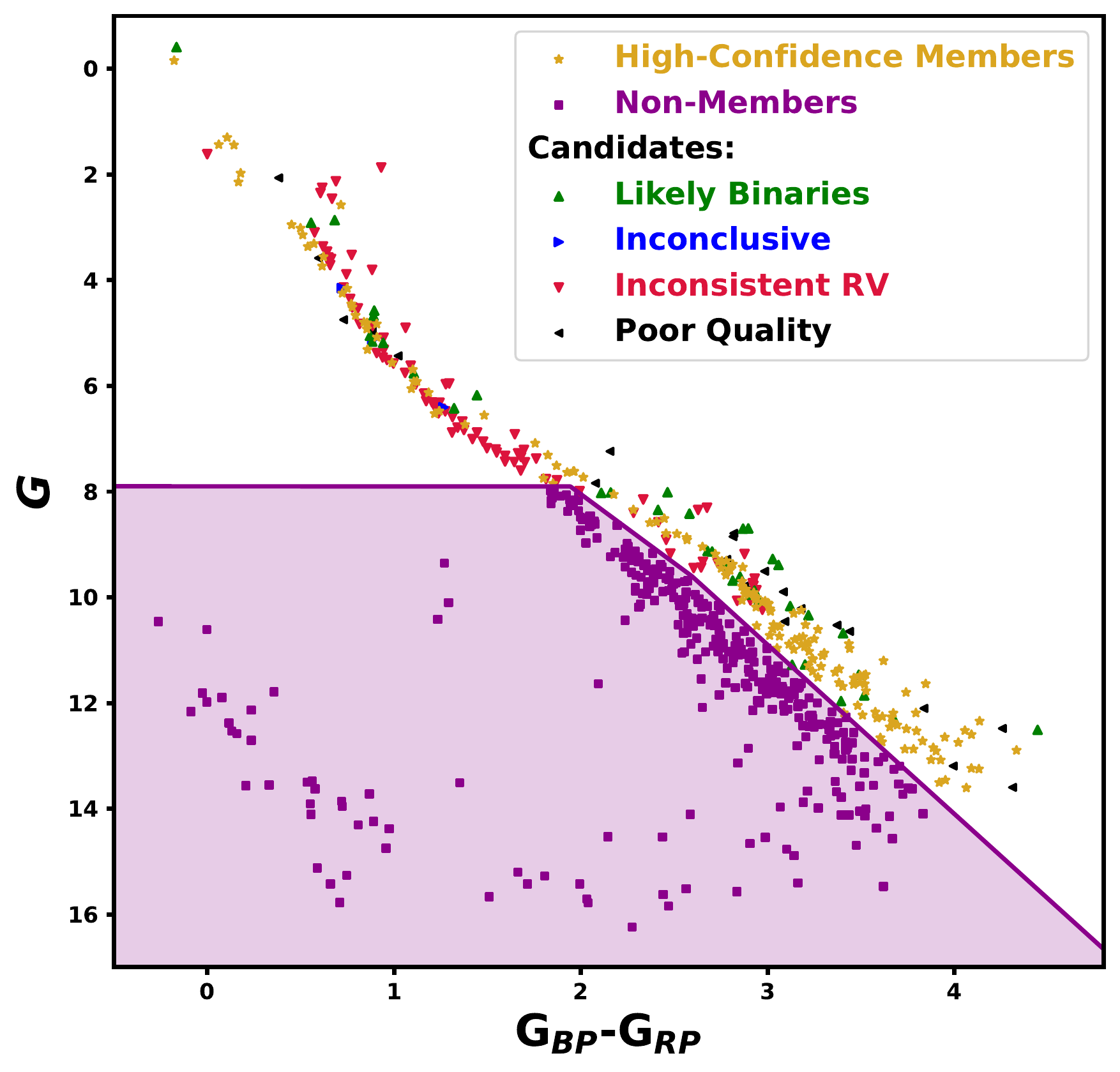}\hfill
\caption{CMD of Fornax-Horologium candidates and members. Yellow stars indicate objects that passed all restrictions. Purple squares were removed by our hard photometric cut, shown by the purple shaded region. Red inverted triangles were removed by our 3D velocity cut. Black left-pointing triangles were removed due to photometric or astrometric quality cuts. Blue right-facing triangles are objects on the pre-main sequence without RV coverage, making their membership inconclusive. Green upward-pointing arrows passed all other cuts but were rejected by our RUWE cut and spectroscopic binarity checks; they consist primarily of stars elevated above the pre-main sequence by the contribution from an unseen companion.}
\label{fig:Mem_CMD}
\end{figure}

In this section we describe the cuts applied to our candidate list for Fornax-Horologium, both to identify members and to perform dynamical analyses on these populations. We compile the results of our stellar selection in Figure \ref{fig:Mem_CMD}, showing which stars survived our cuts, and the reasons that stars were removed. The restrictions we apply fall into three general categories: photometric youth, velocity, and general quality. Only the photometric youth cut is  needed for coherent results in all cases, while the other cuts are necessary for some tasks, but not others. Here we both describe the methods employed to produce these cuts, and the cases in which they must be used. 

\subsection{Photometric Selection} \label{sec:photocut}

Young stars can be readily identified by their elevated luminosities compared to older stars, placing them on the pre-main sequence, which is higher on a color-magnitude diagram than the field main sequence. Fornax-Horologium is no different, showing two distinct sequences in the color-magnitude diagram (Figure \ref{fig:Mem_CMD}). However, due to its age (38.5 Myr, estimated from SPYGLASS-I), Fornax-Horologium is near the upper edge range of ages where our detection of populations is reliable, making it harder to reliably separate from the field through comparison with SPYGLASS stellar population models. As a result of this weaker applicability to the SPYGLASS models, we select photometrically distinct members of the population using a linear cut along the sparsely populated gap between the two sequences rather than using the Bayesian membership approach used for CFN in SPYGLASS-II. This cut and the stars it removed are displayed in Figure \ref{fig:Mem_CMD} (purple shaded region). It was applied to all stars with an absolute magnitude $G>7.9$, roughly corresponding to the area just below the pre-main sequence turn-on. Stars that failed this cut can be interpreted as having photometric properties inconsistent with youth. 

\subsection{RV Selection} \label{sec:rvselect}

Stars that formed together retain those motions from formation, so having a complete 3D velocity vector consistent with the proposed parent association is a strong indicator of membership. Between observations at LCO and the literature radial velocity data, combined with proper motion measurements from Gaia, all possible members with ambiguous photometric ages can have their membership assessed using 3D motions. As in SPYGLASS-II, we approximated the velocity distribution of Fornax-Horologium to be spherical in UVW space, using the projected radius of the transverse velocity distribution as the maximum distance between genuine members and the centre of motion for the association in UVW space. For Fornax-Horologium, we computed a projected radius of 6.9 km s$^{-1}$, and stars within that distance of the association's median UVW value, which is (U,V,W)=(-12.72, -21.61, -4.54), were taken as members.

\begin{figure}
\centering
\includegraphics[width=8.2cm]{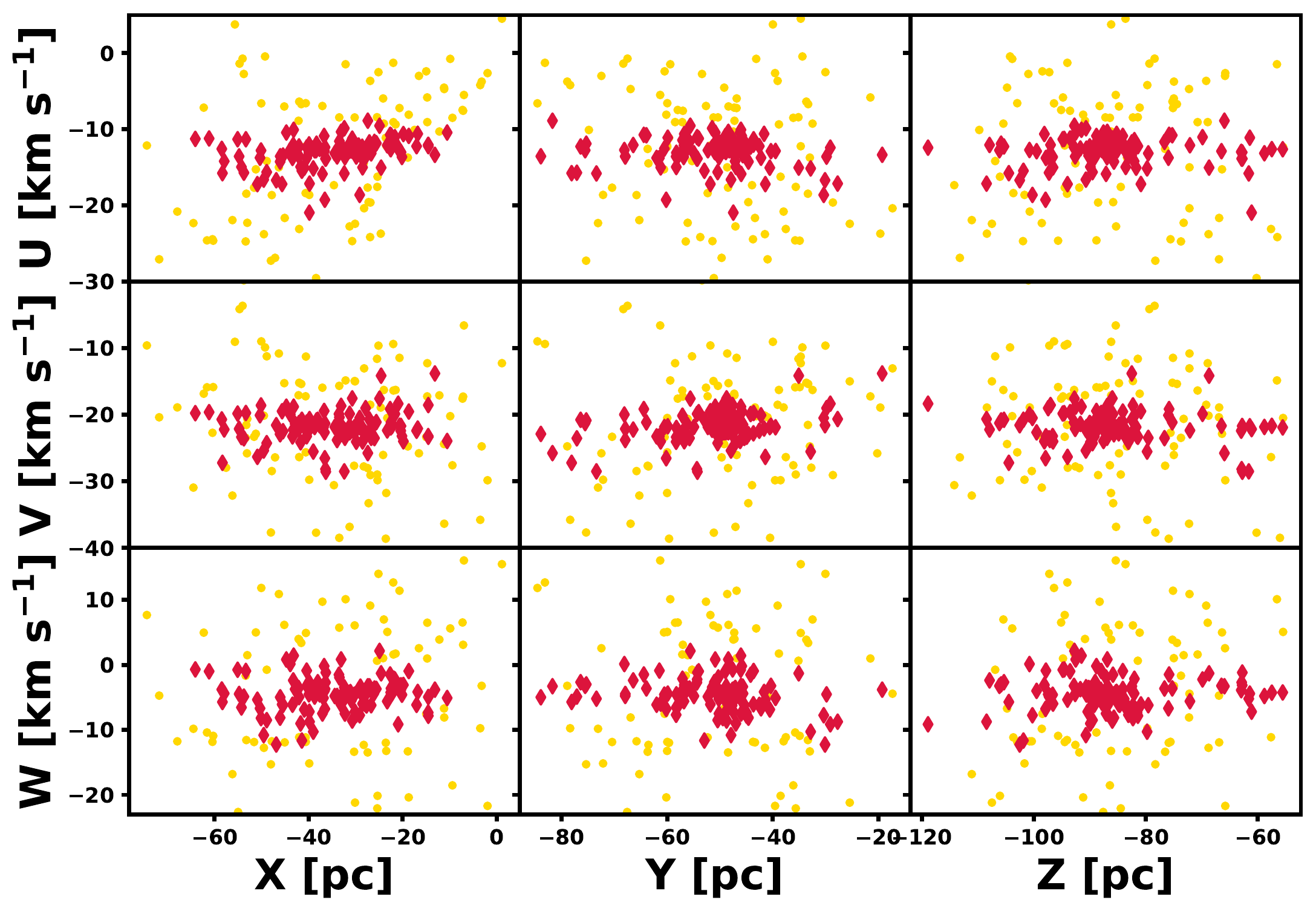}\hfill
\caption{Distribution of the three velocity axes within our updated Fornax-Horologium candidate member sample, plotted against the three spatial axes. Stars marked as red diamonds passed our RV cut, while stars marked by smaller yellow dots did not.}
\label{fig:UVWcuts}
\end{figure}

The resulting velocity selection is shown in Figure \ref{fig:UVWcuts}, where stars that survived are shown in red, and stars that did not are shown in yellow. There is some spatial dependence to the UVW velocities, but its magnitude is smaller than the radius scale of our UVW search, allowing the spatially dependent patterns to the association in UVW space to reliably be separated from the background. The results by star are also presented using a flag in Table \ref{tab:australproperties}, with objects in the external populations being marked as kinematic members, to reflect the fact that these objects have been identified as genuine by our external sources. 

Member identification using radial velocity is complicated by the presence of common-motion interlopers and binaries with high internal velocities. Binaries, in particular, dramatically raise the internal velocities within a system and have the dual effects of causing false negative membership assessments, while also driving up the association's radius in transverse velocity space, which allows field interlopers to get in. While these complications make an RV membership inconclusive, it nonetheless provides a powerful method to limit the sample to some of the more reliable prospective members of the association. 

\subsection{Binaries}

Due to the velocity anomalies induced by the presence of a companion, members of binary systems can produce velocity measurements inconsistent with the velocity barycenter that the system acquired from formation. Stars with orbital velocities exceeding about 7 km s$^{-1}$ will therefore be regularly misidentified as non-members, and stars with more modest velocity anomalies will be rendered unsuitable for use in dynamical studies. Binaries must therefore be identified, both to understand our sample's completeness and to vet the data before they are used to assess the motions of Fornax-Horologium and the stars within.

To locate binaries we used the dual-metric approach from SPYGLASS-II to separately identify resolved and likely unresolved systems. To identify resolved companions, we used a Gaia search, identifying a star as part of a binary system if it has a companion within 10000 AU in the plane of the sky at its measured Gaia distance, provided that it also has a proper motion $\Delta \mu < 5$ mas yr$^{-1}$ and a parallax with $\frac{\Delta \pi}{\pi} < 0.2$. A summary of the likely binaries detected is provided in Appendix \ref{app:bin}.

For identifying likely unresolved binaries, we used the renormalized unit weight error (RUWE), a quality measurement from Gaia of the deviation from a single-solution astrometric model, which can be used to identify the deviations imposed by the presence of a binary companion. Following \citet{Bryson20}, we take stars with RUWE$>$1.2 to be likely binaries. While a more restrictive cut of RUWE$>$ 1.1 will remove nearly all binaries from a sample, it is most useful in more contaminated populations where there are enough stars to maintain a robust sample after the strict cut. For our purposes, a cut of RUWE$>$1.2 is sufficient. 

\subsection{General Quality Cuts}

Stars that failed the photometric and astrometric quality cuts used in this work, and described in SPYGLASS-II, must also be considered for exclusion. These stars have potentially spurious photometric and astrometric solutions, making their velocity or photometric solutions potentially spurious. However, they should still be included in census studies, as a certain number of \textit{bona fide} members will fail these quality cuts, making their inclusion necessary for a complete sample. Conversely, dynamical studies are likely to be adversely affected by the presence of poor astrometric quality, in the same way that poor photometry may produce spurious results in isochronal age estimates. For this reason, we excluded objects that failed photometric or astrometric quality flags from our later analyses, while keeping them for our census studies.

\subsection{Stellar Selection Summary} \label{sec:sss}

Following SPYGLASS-II, we employed two separate sets of cuts for different purposes: one purely photometric cut for studying the stellar populations in FH and their potential substructure, and a more holistic method for establishing a population of probable FH members. Our sample for investigating the population demographics and structure simply consists of all stars except for those that failed the photometric selection cut described in Section \ref{sec:photocut}. This cut left 329 candidate members in our sample, and this is the complete population of credible Fornax-Horologium members that is presented in our member tables. We applied this same cut to the full set of 514 literature Tuc-Hor, Columba, and Carina members that comprises our basic data set used for structural and population analyses, removing 24 in that sample. After combining all of our samples and removing 8 duplicates found in both our Fornax-Horologium sample and our literature-derived catalog of Columba members, our sample size across all Austral sub-associations is 811.  

Our probable member sample also closely follows the choices made in SPYGLASS-II. We first removed stars with radial velocities inconsistent with Fornax-Horologium, following the RV selection metric described in Section \ref{sec:rvselect}, which removes stars with UVW velocities more than 6.9 km s$^{-1}$ from the association's median motion. For the remaining stars that did not fail either our photometric youth or our velocity agreement cuts, we provided two routes for remaining in the sample: robust photometric youth or robust velocity agreement. For our robust photometric youth condition, we kept any star that passed our photometric youth cut with $G>7.9$, provided that the star also passed our photometric and astrometric quality cuts. For our robust velocity agreement condition, we kept any star that passed our velocity cut, while also passing the Gaia astrometric quality cut to ensure the reliability of the Gaia transverse component of velocity. 

Finally, we removed all stars with spectroscopic evidence of binarity or RUWE$>1.2$. This cut removed most unresolved binaries, along with the complications associated with handling them. This choice is important for a variety of studies, since the presence of an unresolved companion can introduce RV anomalies, and also change the flux received from a system. The latter has the effect of raising the system on the CMD, and increasing the continuum relative to a spectral line, producing underestimates of spectral line equivalent widths. The more restrictive cut (removing stars with RUWE$>$1.1) is excessive for most purposes, which is why we did not apply it here. However, it is useful for our isochronal age estimates, where the stellar samples are generally larger, increasing both the resilience of the sample to strict inclusion vetting as well as the likelihood that binaries with less extreme RUWEs remain in any given sample. We did not remove unresolved binaries at this stage, since the presence of a companion only significantly influences radial velocity, which is only essential for dynamical studies. This cut is critical for ensuring the purity of our dynamical studies, however.  

In the external populations, including Tuc-Hor, Columba, and Carina, we assumed all recorded members to be genuine co-moving members, so our cuts only pertain to maintaining the quality of the sample we employ. For our high-confidence sample we therefore removed all stars with failed astrometric quality flags, as well as objects with RUWE$>$1.2, following the same considerations made in selecting our Fornax-Horologium RUWE cut. A further requirement that the photometric quality flag is also passed is necessary for tasks involving magnitudes, such as isochrone fitting and studying the lithium sequence, but it is not necessary for traceback. Similarly, the removal of resolved binaries is necessary for traceback, but nothing else. 

\input{AUSTRAL_MEM}

\section{Results} \label{sec:results}

\subsection{Substructure} \label{sec:subclustering}

\begin{figure*}
\centering
\includegraphics[width=17.5cm]{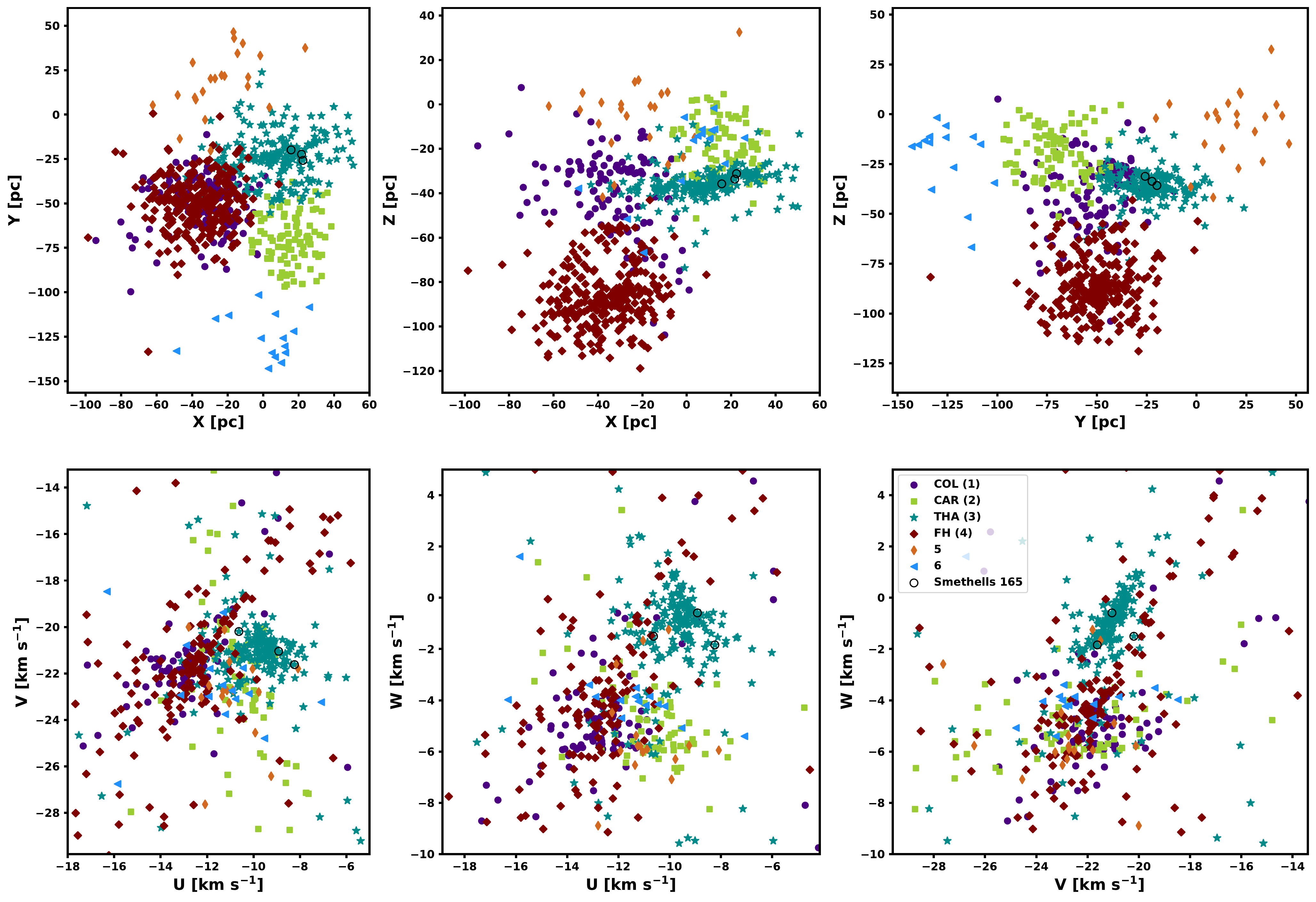}\hfill
\caption{Distribution of stars in Fornax-Horologium and the other Austral sub-associations re-clustered using HDBSCAN, shown in XYZ galactic Cartesian coordinates on the top row, and UVW velocity coordinates on the bottom row. The three proposed Smethells 165 association members are enclosed within black circles, and all are included in and fully consistent with Tuc-Hor. The symbols assigned to members of each subregion within the Austral complex are marked in the legend in the bottom-right panel.}
\label{fig:australspatial}
\end{figure*}

The combination of our new catalog for Fornax-Horologium with our literature compilation of other Austral sub-associations produces a contiguous distribution of stars in spatial and velocity coordinates. Fornax-Horologium firmly connects to the rest of the Austral Complex via Columba, with 8 members shared between our FH list and the literature Columba list we use here. This new and contiguous view of the Austral Complex allows for re-clustering of the entire sample to assess whether the current division of the association into FH, Tuc-Hor, Columba, and Carina is appropriate, and whether additional substructure may be present.

The proximity of this complex skews the transverse velocities due to geometric effects, complicating the 2D view of the association. Meanwhile, we have significantly expanded our 3D view of velocities in the Austral association using both observations and literature, covering approximately three-quarters of the combined Austral sample. We therefore chose to cluster in 6D space-velocity coordinates, using the full 3D UVW velocity vector in place of the 2D transverse velocity anomaly parameter used in SPYGLASS-II, alongside the XYZ galactic coordinates for the spatial dimensions. We used the HDBSCAN hierarchical clustering algorithm to cluster the resulting sample, making use of the ``EOM'' (excess of mass) clustering method, which identifies the most ``persistent'' clusters: those identified over the widest range of clustering scales. Velocity axes were multiplied by a scaling factor $c = 6$ pc/km s$^{-1}$, which ensures that the scales of the spatial and velocity coordinates are comparable and conducive to clustering together. This is the same value used to convert between velocity and spatial axes in SPYGLASS-II. For HDBSCAN clustering, we set {\tt min$\_$samples} and {\tt min$\_$cluster$\_$size} to 6, reflecting slightly more sensitive clustering compared to SPYGLASS-II, with the intention of compensating for the sensitivity loss from the necessary exclusion of members from our 6D clustering due to their lack of a radial velocity vector. We also attempted clustering on the same sample using the HDBSCAN leaf clustering method instead of EOM, which was used in SPYGLASS-I to identify further substructure within subgroups. This choice resulted in the fragmentation of Tuc-Hor as well as Columba. However, upon visual inspection we found these results to be quite tenuous and, given the large uncertainties of some of the radial velocity measurements used to separate them, we conclude that the identification of further substructures is inconclusive, and therefore only report EOM clusters.

The resulting clustering produces central cores for likely subgroups, but many stars are not assigned to a subgroup, either because their lack of an RV excluded them from analysis, or due to their assignment to the background, a possibility enabled by HDBSCAN's optimization for a contaminated environment. Since our input population is presumed to be composed mostly of genuine members, we can widen our coverage of each group by assigning outlying stars to the nearest group. Not all stars will have clear membership in one group over another, leaving a possibility of incorrect group assignment in edge cases, however we can nonetheless assign stars to the parent group that is most likely given our current data set. We use the 5-D galactic XYZ/transverse velocity anomaly distance metric used in SPYGLASS-II to compute which group is nearest, which allows us to include the numerous stars that lack a radial velocity measurement. We used a value of $c = 12$ pc/km s$^{-1}$ to convert between the velocity and spatial coordinates, as in SPYGLASS-II. This choice weights velocity somewhat higher, therefore providing a selection better suited to these outer regions where spatial assignment is more ambiguous, given that our core regions are most initially differentiated in spatial coordinates. More information on the choices made for our clustering process can be found in SPYGLASS-II.

The resulting clusters are shown in spatial and velocity coordinates in Figure \ref{fig:australspatial}. Their distributions are more coherent in spatial coordinates than in velocity coordinates, due primarily to the more consistent precision of the spatial coordinates compared to the velocities, especially the radial velocities. This is reflected in the clustering results, which tend to adhere more closely to the spatial distribution than to the velocity distribution. This has the possible downside that stars which escaped their original group and are now superimposed in front of another may be misassigned due to limitations to the current radial velocity coverage and the geometric effects on projected sky velocities at these close distances. This motivates the further expansion of the radial velocity coverage in the region, to make velocity clusters more resolvable. Nonetheless, all six groups that result produce a reasonably coherent core in both spatial and velocity coordinates. The assignment of stars to the subgroups is included in Table \ref{tab:australproperties}. 

\begin{figure*}
\centering
\includegraphics[width=17cm]{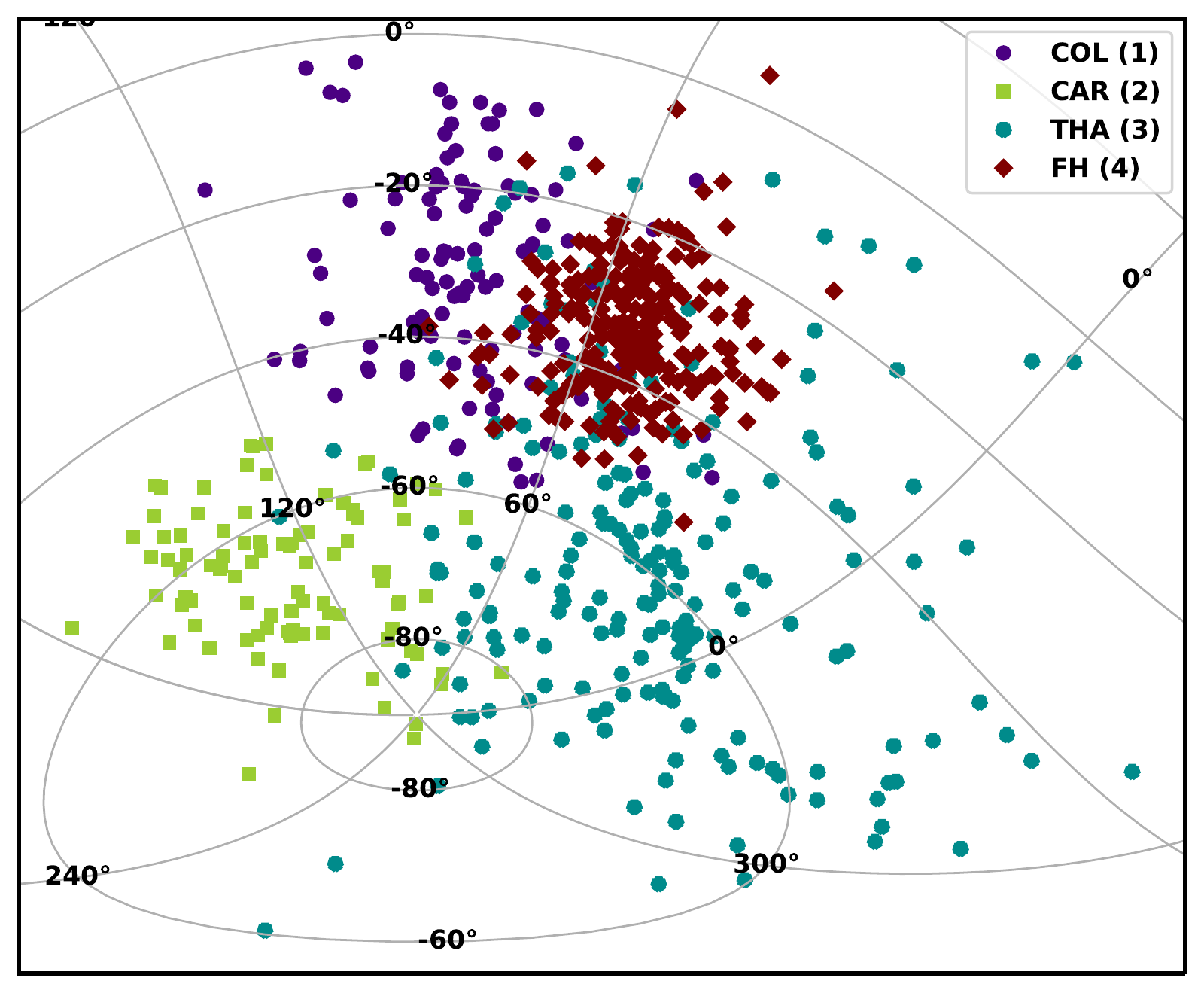}\hfill
\caption{Members of the four well-established Austral sub-associations of FH, Tuc-Hor, Columba and Carina in RA/Dec sky coordinates.}
\label{fig:australsky}
\end{figure*} 

The four well-established Austral sub-associations (FH, Tuc-Hor, Columba, and Carina) all emerge from our clustering in forms similar to those that have previously been established. The extent of these core Austral populations in the plane of the sky is shown in Figure \ref{fig:australsky}. There are also two new structures that our clustering identifies as possible new subgroups. First, our clustering assigns the northernmost section of the Austral Complex to a new subgroup provisionally referred to as Group 5, which consists mainly of previously identified Columba members north of the celestial equator. The second new region, which we call Group 6, is approximately 50 pc behind the main body of Carina in the plane of the sky. It is defined using only stars from \citet{Torres08}, all but three of which were identified as Carina members. We discuss whether these groups are real and whether they connect to the Austral association in Sections \ref{sec:g5} and \ref{sec:g6}, and conclude that Group 5 has insufficient evidence of youth, and Group 6 is mostly composed of a subcluster of Sco-Cen, rather than being a new component of the Austral Complex. As a result, we do not include these regions in our age assessments and de-emphasize them in the subsequent reconstruction of the Austral association's formation. 

Overall, the positions of the subgroups in spatial coordinates are fairly continuous and overlapping, with the possible exception of Group 6. However, there is a visible dichotomy in velocity between Tuc-Hor and all other subgroups, separated by about 4 km s$^{-1}$ on average. This may hint to a more distinct dynamical history for that region.

\input{SG_PROPS}

\subsubsection{Smethells 165} \label{sec:smth165}

The Smethells 165 association, which was proposed as a dynamically distinct region that is spatially overlapping with Tuc-Hor \citep{Higashio22}, was notably absent from our clustering results. All three proposed members were grouped under Tuc-Hor by our clustering, without any hint of substructure associated with them. While we were able to reconstruct the velocity vectors used to define the association in \citet{Higashio22} using Gaia DR2 RVs, which do appear distinct from the rest of Tuc-Hor, we found that the high-quality non-Gaia RVs in our final sample were very different. With our improved velocity database, we found that all three proposed members were clearly consistent with the distribution of Tuc-Hor velocities. This result strongly suggests that Smethells 165 does not form a separate association, but instead emerges as an artefact from RV anomalies in the Gaia data.

\subsubsection{Is Group 5 Real?}\label{sec:g5}

Group 5 is the sparsest Austral subgroup that we identified. Its tenuous nature is reflected in its CMD and lithium sequence, both of which lack much coverage in the sections of the sequences that significantly leverage the age solutions. Removing our quality vetting for isochronal analyses does reveal a sequence that continues to dimmer magnitudes, but all of these stars barely pass our photometric youth cut, casting doubt on whether this group is actually young. It does have a reasonably consistent set of velocities, which at least indicates the presence of some common population, and the currently weak coverage of the region is perhaps insufficient to conclude that the group is certainly older, so we do not discount the possibility of this being a genuine new subgroup. These results motivate a search for phase-space neighbors prior to concluding whether it should be grouped under the Austral association. However, given the currently available data there is no clear evidence that Group 5 is nearly as young as the other Austral subregions. 

\subsubsection{What is Group 6?} \label{sec:g6}

Group 6 members form a majority of the \citet{Torres08} Carina members, while our definition of Carina is much more closely aligned with the distribution noted in \citet{Gagne18BXIII}. The notable spatial gap between these \citet{Torres08} Carina members and our main body prompted us to check for association assignments for members of Group 6 in SPYGLASS-I. The result was that 9 of its 14 members were found to be grouped under the Sco-Cen complex, with 6 of those assigned specifically to the Platais 8 cluster within Sco-Cen. The velocity core of Group 6 overlaps fully with Platais 8, so we conclude that the group consists of that cluster in addition to some outlying neighbors. By extension, we can also determine that the \citet{Torres08} definition of Carina was centered on Platais 8, not the modern extent of Carina, although some genuine Carina members were included. Platais 8 has a very similar age to components of the Austral association, with a SPYGLASS-I age of 37 Myr compared to the 38.5 Myr SPYGLASS-I age of FH. An age in this range is supported by the literature lithium coverage we collected in this region, which also agrees with the lithium sequence shown elsewhere in the Austral complex. While the notable spatial gap between Carina and Platais 8 suggests that the latter should not be included as part of the Austral association, it is nonetheless useful to include during traceback to investigate possible links between the Austral association and Sco-Cen.  

\subsection{Ages}

In this section, we assemble age solutions from all sources available to us. Through our extensive coverage of the Austral Complex, ages can be assessed using isochronal, dynamical, asteroseismic, and lithium depletion methods, before being synthesized into combined ages most consistent with all data available. We compile the results for each one in Table \ref{tab:ages}, where we also include a combined age for each sub-association, which is produced in Section \ref{sec:ageadopted}.

\input{AGES}

\subsubsection{Isochronal Age}

\begin{figure*}
\centering
\includegraphics[width=16.5cm]{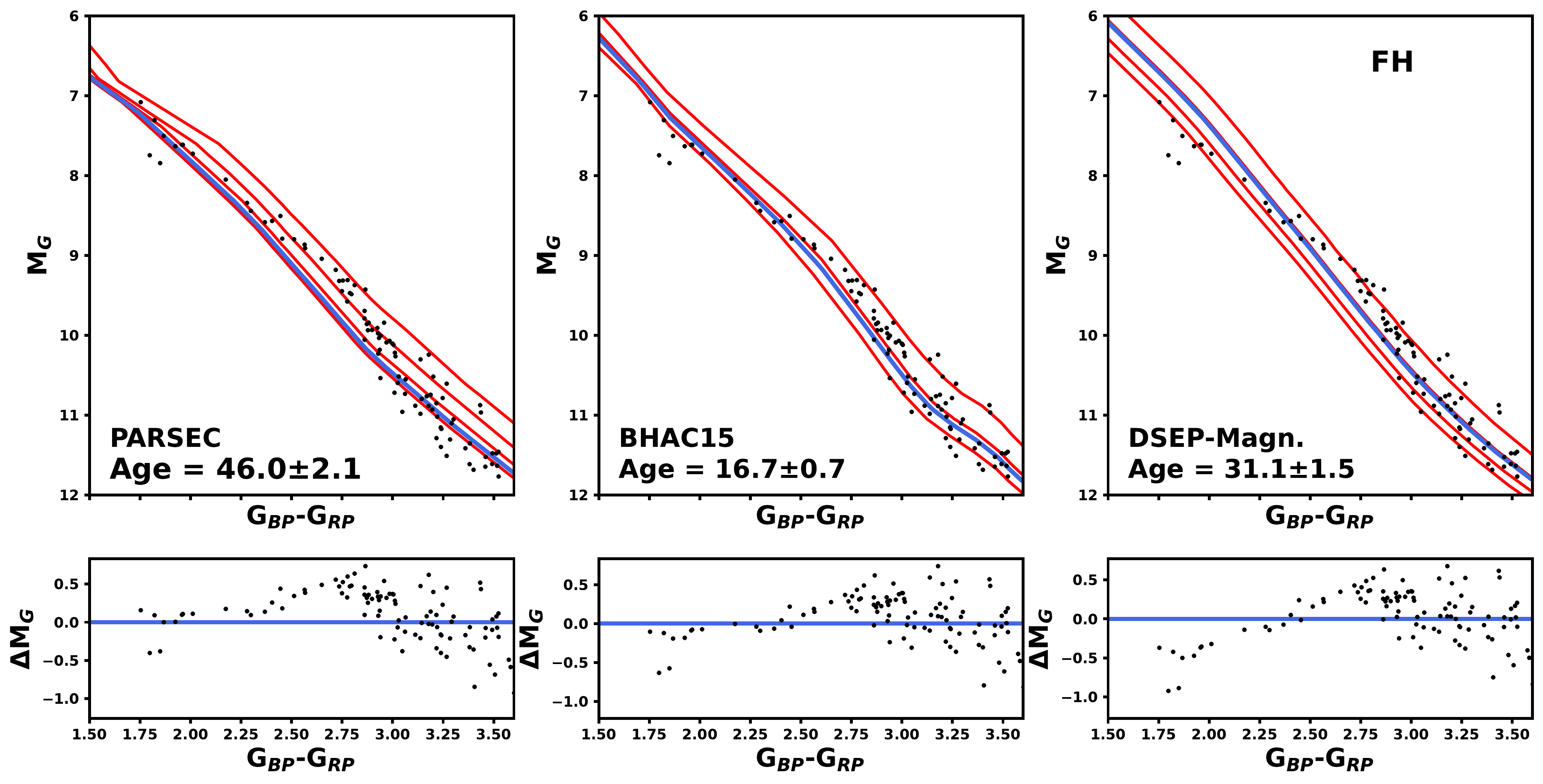}\hfill
\caption{The Isochronal Age fit for the Fornax-Horologium Association, as defined by our re-clustering in Section \ref{sec:subclustering}. The top row shows the color-magnitude diagram of the members included in the group. Stars included in the fitting are shown as black dots, and the best fit result is in blue. We also provide a range of isochrones for comparison which, from top to bottom, are: 20, 30, 40, and 50 Myr for PARSEC; 10, 15, and 20 Myr for BHAC15; and 20, 30, 40, and 50 Myr for DSEP-Magnetic. The bottom row provides residuals between the best fit model and photometry. Fits for Carina, Columba, and Tuc-Hor are provided in the online-only version of this figure.}
\label{fig:isoagefits}
\end{figure*}

Most age estimates provided to date have been isochronal, made possible by the general accessibility of photometry relative to spectral observations, especially with the recent Gaia data releases \citep[e.g.,][]{Zuckerman19, Galli21}. Isochronal ages are a critical tool in our measurement of ages, but the resulting ages can vary significantly between different models, especially on the pre-main sequence \citep[e.g.,][]{Herczeg15}. This makes isochronal ages an excellent option for age comparison between stellar populations of the same origin, but they can be significantly less reliable in establishing accurate absolute ages compared to other methods, such as lithium depletion or dynamical ages \edit1{\citep[e.g.,][]{Binks22}}. Regardless, with our robust membership diagnostics for stars in the Fornax-Horologium association, together with the literature samples for the other Austral sub-associations, we can provide isochronal ages throughout the association. 

Our isochrone fitting followed the method outlined in SPYGLASS-I (Section 3.5) and later used in SPYGLASS-II, using least-squares optimization on a photometric sample gathered from Gaia EDR3, restricted to a range of $G_{BP}-G_{RP}$ which covers most of the pre-main sequence. We calculated fits according to the definitions of groups assigned in Section \ref{sec:subclustering}, clustering on the high-confidence version of our member samples that are described in Section \ref{sec:sss}, with the additional requirement RUWE$<$1.2. This is a looser requirement than used in SPYGLASS-II, and was chosen because that paper's restriction to RUWE$<$1.1 removes entire sections of some Austral CMD sequences (in the case of Tuc-Hor, it results in a gap spanning two magnitudes in $G$ for the K and M stars). With the models occasionally undershooting or overshooting the stellar sequence, it is important that the coverage of stars across the sequence is reasonably uniform, such that deviations of the model have uniform effects when fitting each sequence. We found these looser constraints populated these sequences more fully, while not adding an obvious binary sequence. We calculated fits for FH, Tuc-Hor, Columba, and Carina, but excluding the tenuous new groups presented in our clustering results, which lack significant coverage on the pre-main sequence. 

We show the resulting isochrone fits in Figure \ref{fig:isoagefits}. The same three isochrone grids used in SPYGLASS-II were assembled for fitting: PARSEC v1.2S \citep{PARSECChen15}, BHAC15 \citep{BHAC15}, and DSEP-Magnetic \citep{Feiden16}. The grid for the \citet{PARSECChen15} and \citet{BHAC15} isochrones was limited to $1.2<G_{BP}-G_{RP}<4$, while the DSEP-Magnetic isochrones, which do not go quite as far to the red, were restricted further to $G_{BP}-G_{RP}<3.6$. We compile the resulting age solutions in Table \ref{tab:ages}. 

The results for Fornax-Horologium and its companions show a fairly consistent story. All three models show Tuc-Hor to be the oldest, and all but DSEP-Magnetic show Carina as the youngest. Columba aligns quite closely with Carina, with an age younger than found for Carina using DSEP-magnetic models. FH is slightly older than Carina and Columba, but still much closer to them in age than to Tuc-Hor. Age similarities between FH and Columba are not surprising, given that their definitions in the literature and this work overlap along FH's northern edge. While many publications have found no significant age differences between these populations \citep[e.g.,][]{Bell15, Zuckerman19, Schneider19}, the group most consistently identified as old is Tuc-Hor, with age solutions consistently near to or older than 40 Myr \citep[e.g.,][]{Kraus14,Bell15,Galli21}. Carina has seen age solutions as young as 22 Myr \citep{Schneider19} and even 13 Myr \citep{Booth21}, and the $\chi^{1}$ For region of FH has seen published ages closer to 30 Myr \citep{Mamajek16, Galli21}. The sequence we observe therefore roughly reflects the overall sequence of ages reported in the literature.

\subsubsection{Lithium Depletion} \label{sec:lithdep}

Lithium depletion measurements have generated some of the most reliable ages for the Austral Association to date. Unlike for isochronal ages, models generally agree on the absolute age scales of lithium depletion, making them more reliable for assessing absolute age scales \citep{Binks14}. This makes them a valuable supplement to the isochronal ages, since while the latter are sensitive to relative age differences but unreliable for absolute values, lithium ages are reliable in absolute terms but often require deep stellar samples to be able to resolve differences between subgroups (such as coverage of the Lithium Depletion Boundary (LDB)). While many stars in the Austral association have lithium measurements, only Tuc-Hor has significant coverage that extends to the LDB. Our new observations at LCO were limited to $m_G \la 13$, which does not enable coverage of the LDB, however it does provide significant lithium coverage at higher masses, where lithium depletion occurs slowly through convection into lithium-burning regions deeper in the star.  

\begin{figure}
\centering
\includegraphics[width=8.0cm]{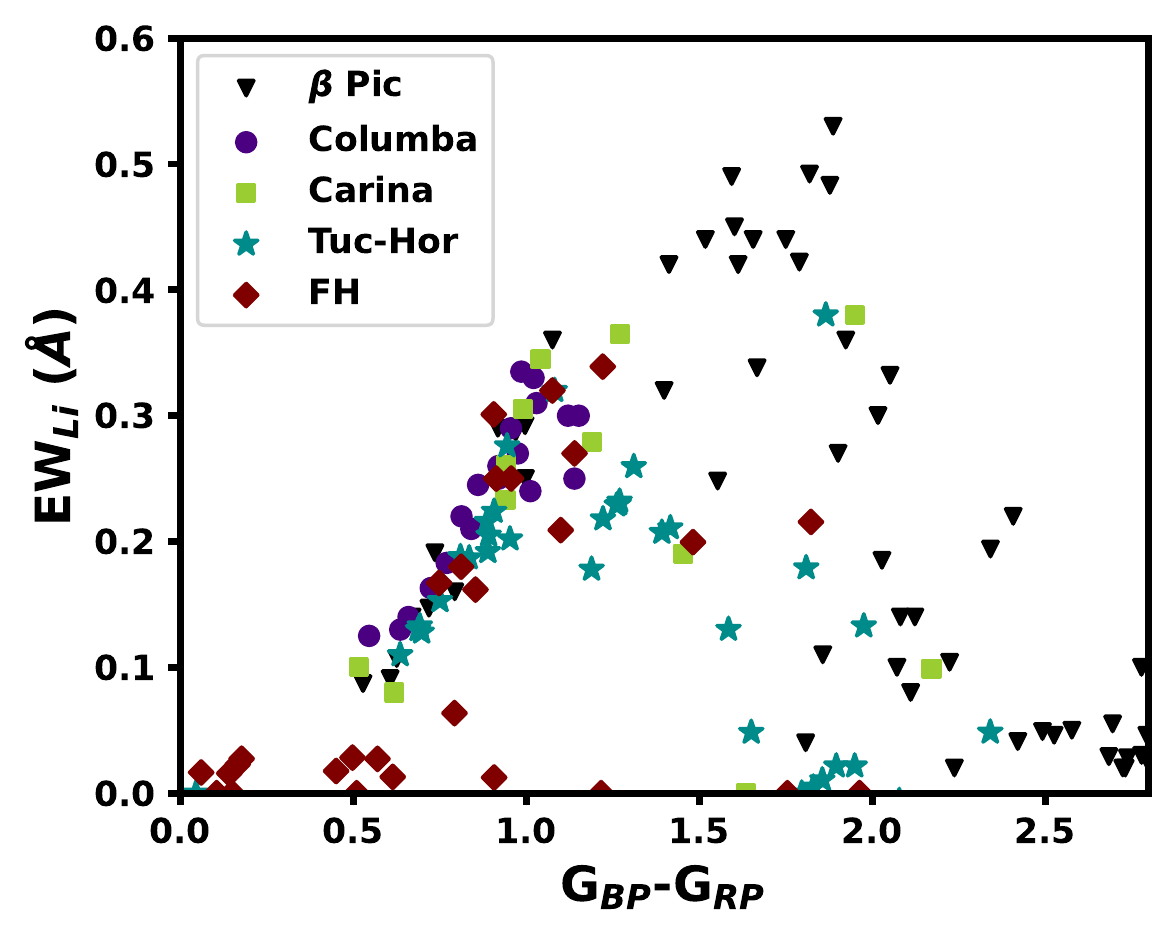}\hfill
\caption{Lithium sequence of stars in the four well-established sub-associations of the Austral complex. Markers are the same as in Figure \ref{fig:australspatial} and are also shown in the legend. We also include $\beta$ Pic ($\sim$22 Myr) for reference, marked as black inverted triangles.}
\label{fig:lisequence}
\end{figure}

The lithium sequence in the Austral Complex is shown in Figure \ref{fig:lisequence}, derived from the probable member sample which excludes likely unresolved binaries and RV outliers. We include $\beta$ Pic members for reference. We see that across all Austral subregions, lithium sequences are largely in agreement with one another. Carina is \edit1{the} only subregion where Li EWs are credibly different, with members generally elevated above those of other Austral sub-associations. The limited number of stars present motivates caution in interpreting these differences and not all members have elevated Li EWs, however nearly all stars with outlying measurements belong to Carina. Only one star with lithium clearly enriched above the Austral Complex's main sequence exists outside of Carina and it was originally in our literature Carina sample prior to re-clustering. This demonstrates that the Lithium-enriched sequence could plausibly be a feature unique to Carina. \citet{Schneider19} used the stars on Carina's young lithium sequence to justify an age consistent with $\beta$ Pic at $\sim$22 Myr \citep{Binks14, Mamajek14}, a result further strengthened by their inclusion of two stars which constrain the age at the LDB, which were excluded by our quality cuts. While the lithium sequence of Carina does provide evidence for younger stars being present in the region, the result does not discount the possibility of older populations being mixed in, or even providing a significant contribution to the overall population.

Aside from Carina, all sequences appear largely indistinguishable from one another. However, our ability to resolve differences between sequences is considerably weakened by the lack of Li measurements for stars with $1.2<G_{BP}-G_{RP}<2.0$ where these abundances change most quickly for this age range, with the exception of Tuc-Hor, which has much deeper coverage through \citet{Kraus14}. The results nonetheless show broad consistency with a 30--50 Myr age range, clearly older than $\beta$ Pic at $\sim$22 Myr \citep{Binks14, Mamajek14}, but consistent with the Tuc-Hor sequence of \citet{Kraus14} and the 36--45 Myr age range suggested by that publication. Any further information from lithium depletion on the formation sequence of these regions is therefore largely inconclusive, motivating the further expansion of this coverage through future projects, especially to the LDB.

\subsubsection{Dynamical Ages}

\begin{figure}
\centering
\includegraphics[width=8.2cm]{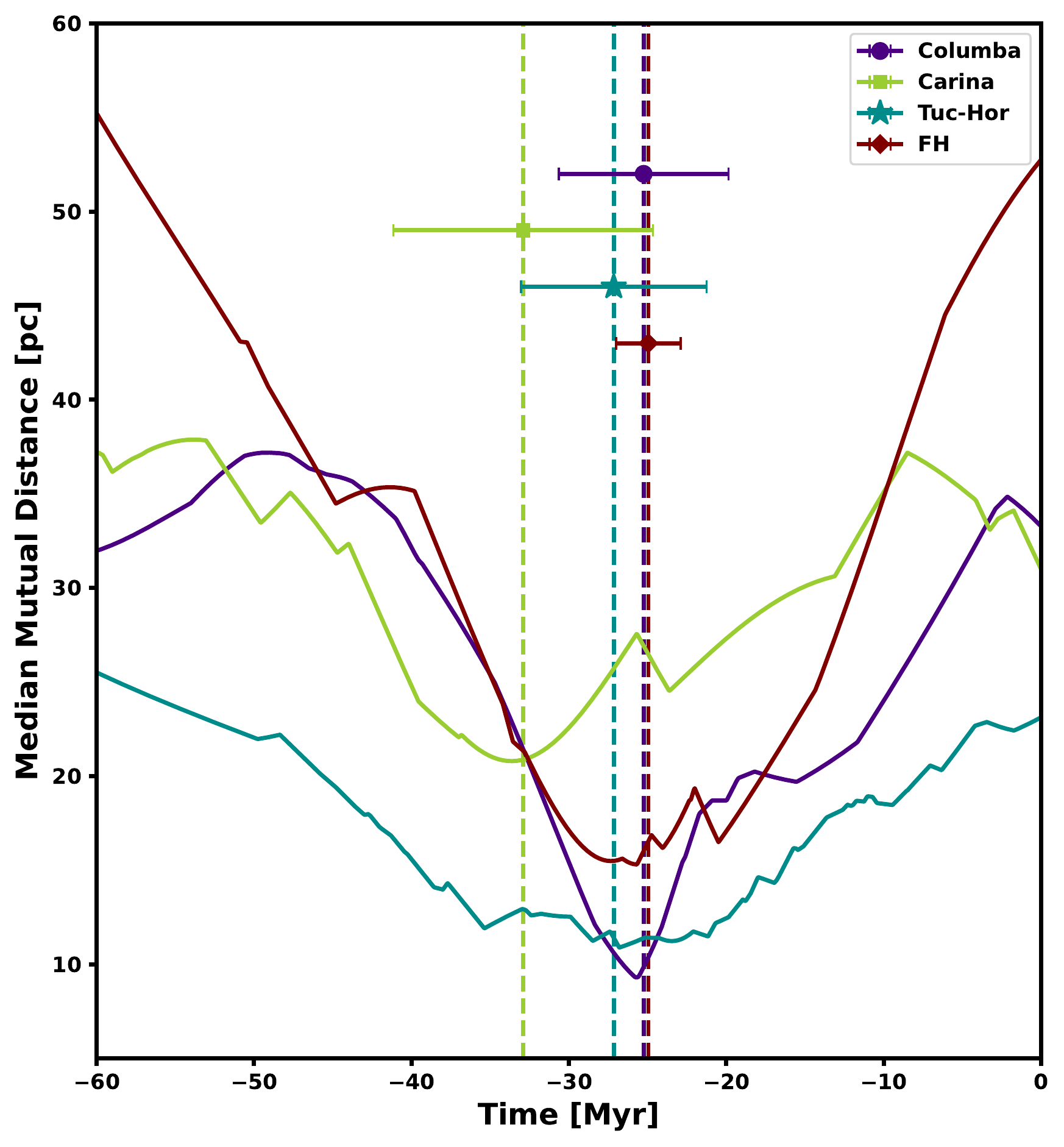}\hfill
\caption{Dynamical ages for FH, Tuc-Hor, Columba, and Carina, as defined in our re-clustered sample. The best fit age is marked by the vertical lines and markers, with the markers matching what is used in Figure \ref{fig:australspatial}. The curves show the median mutual distance between stars as a subgroup of time, with dynamical ages correlating with the minima of those curves.}
\label{fig:dynmages}
\end{figure}

As in SPYGLASS-II, we computed dynamical ages by searching for a moment of closest configuration in the histories of Fornax-Horologium and its companion associations. First, we applied a few additional restrictions to our member samples to ensure the quality of our populations for traceback. Across Fornax-Horologium, Columba, Tuc-Hor, and Carina (as defined by our re-clustering in Section \ref{sec:subclustering}), we removed all stars with resolved companions, as well as stars with $\sigma_{RV}>0.5$ km s$^{-1}$. This velocity quality cut is more restrictive than the choice made in SPYGLASS-II, motivated by the higher ages expected in the Austral complex compared to CFN, which require better velocities to achieve the same level of positional accuracy around the time of formation. We also removed stars represented by Gaia RVs, since previous experimentation in SPYGLASS-II demonstrated that the Gaia RVs provide much less consistent traceback compared to ground-based observations.  We found that the echelle spectrographs at the McDonald Observatory 2.7m Telescope and LCO gave RVs that consistently traced back to a tighter past configuration. 

Due to the diversity of sources our RVs were drawn from, we also made two additional cuts to remove velocity outliers within individual populations. First, we removed stars outside of the 3-$\sigma$ extent of an ellipsoid, with the extents along the U, V, and W axes set by a 3-$\sigma$ sigma-clipped standard deviation and median of stars along that axis. Our second cut concerns the slight bimodality in the Carina and Tuc-Hor velocity distributions mainly along the V axis, in which an overdense scattering of Tuc-Hor members overlap with the Carina velocity core and vice versa. Ensuring correct subgroup assignment is particularly important for dynamical ages, as if the stars are drawn from multiple populations a dynamical age calculation may recovery a past close approach between subgroups, not a tight past configuration within a group. The conclusion is that these two groups are particularly vulnerable to mutual misidentification, and it is therefore best to remove any stars where membership in the other is a distinct possibility. We therefore exclude Carina members with V$>-22$ km s$^{-1}$, and Tuc-Hor members with V$<-22$ km s$^{-1}$. For both regions, the stars removed only marginally survived the 3-$\sigma$ ellipsoid cut, with the populations removed including only 3 stars in the Carina sample and 2 in the Tuc-Hor sample. With these restricted populations, we used galpy's numerical integration routine with the \texttt{MWPotential2014} Milky Way potential model \citep{Bovy15} to perform dynamical traceback.

For each age step and star, we compute a median distance to all other association members, and for each star we return the time when that median distance is the smallest. This provides an individual sample of the stellar traceback age. The distributions of most of the resulting individual traceback ages tended to be bimodal, containing a peak around zero, likely composed of stars with bad RVs, undetected binary companions, or membership in bound embedded clusters like ${\chi}^1$ For (see Section \ref{sec:fhvir}). To minimize their influence, we removed stars with individual age solutions younger than 10 Myr, which is a limit well below any ages being considered in these regions. The individual traceback ages were then computed again on this downselected sample, with the resulting age computed as the median of these closest approaches to each star's neighbors, with an uncertainty equal to the standard deviation, as in SPYGLASS-II. 

Our age fits are shown in Figure \ref{fig:dynmages}, alongside the median mutual distance between stars in the sample, which provides a visual way of viewing the convergence of stars around the time of their formation. The result is age solutions that heavily overlap within uncertainties. The formation order is not in complete agreement with that of our isochronal age solutions, however the uncertainties are large enough that the sequence presented by isochrones is quite plausible. 

As we later show in Section \ref{sec:fhvir}, dynamical age results may be complicated by gravitationally bound subregions within these populations, such as ${\chi}^1$ For, which represents a plausibly bound core to the Fornax-Horologium association. While a precise assessment of its virial state is beyond the scope of this publication, Tuc-Hor also has a dense central region not unlike what is found in FH, suggesting it may also be subject to non-negligible internal gravitational influence. The influence of internal gravitational potentials on our quality-restricted population is however unclear. Stars ejected from a bound core after formation will have traceback ages younger than the association, making ages in such an environment lower limits to the true value. With the older ages suspected for these populations compared to CFN, there is also much more time for early populations to be blended, potentially removing a lot of our resolution in detecting individual distinct star forming environments. This may embed signals in our data related to close approaches between constituent subgroups during the assembly of these associations \citep[e.g.,][]{Bonnell03,Guszejnov22}, however due to the uncertainty in the historical potential many of these details will be extremely difficult to recover.

\subsubsection{Asteroseismic Ages}
\label{sssec:seismology}

Pulsating A and F stars have recently become established as useful age indicators, especially near the terminal-age main sequence as their nuclear fuel runs out \citep[][and references therein]{aerts2021}, or near the zero-age main sequence, where their pressure-mode pulsations follow regular patterns \citep{beddingetal2020}. In the latter context, ages have been inferred asteroseismically for $\delta$~Scuti stars in Upper Centaurus--Lupus \citep{murphyetal2021a} and Cepheus Far North \citep{Kerr22}, and have helped to constrain the age of the Gaia--Enceladus stream \citep{beddingetal2020}. However, the ages are only as good as the physics of the models, and the accretion physics of pre-main-sequence stars can be very complicated \citep{steindletal2022,steindletal2022NatCo}.

The biggest uncertainty in modelling young stars is rotation, which not only decreases the mean stellar density and thereby the frequency spacing of the modes ($\Delta\nu$), but also affects the frequencies of individual pulsation modes in complicated ways \citep{dicriscienzoetal2008,reese2022}. It also delays the pre-main-sequence contraction. For slow rotators, such as HD\,139614 in UCL, the effects of rotation are small \citep{murphyetal2021a} but for some rapidly rotating stars in the $\sim130$-Myr-old Pleiades cluster the effects are large \citep{murphyetal2022a}. The problem of rotation is not limited to asteroseismology, but also affects the accuracy of main-sequence turn-off ages and isochrones more generally \citep{brandt&huang2015a}. One advantage of asteroseismology is that rotation can be measured when rotational splittings of the pulsation modes are identifiable \citep[e.g.,][]{kurtzetal2014,saioetal2015}, but this is difficult for rapid rotators.

\begin{figure}
\centering
\includegraphics[width=8.0cm]{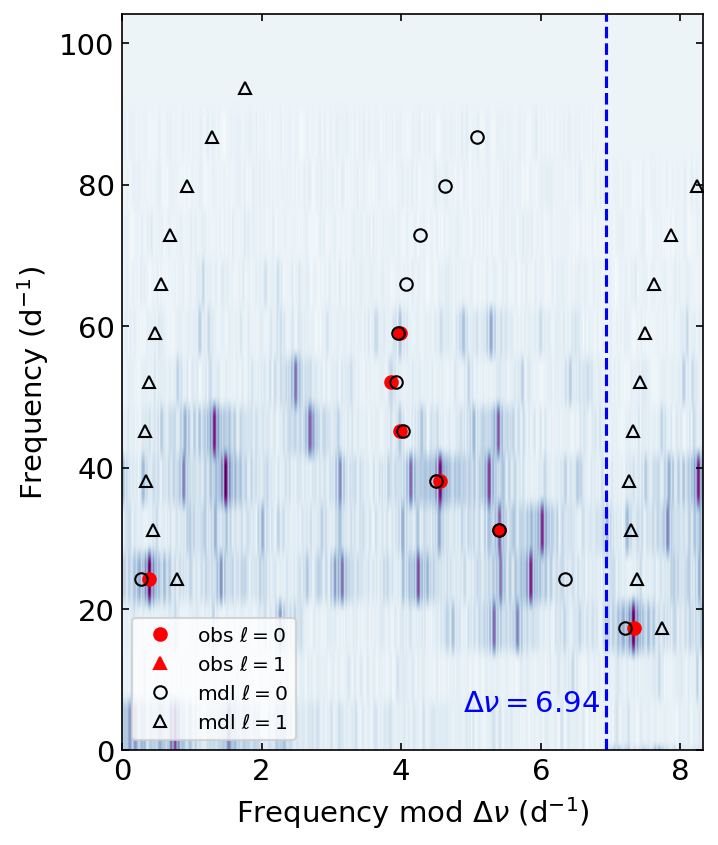}\hfill
\caption{The \'echelle diagram of HD\,21434, where the grayscale shows the Fourier amplitude spectrum after dividing into equal segments of width $\Delta\nu$ and stacking vertically. The vertical dashed line marks $\Delta\nu$ and there is a repeated overlap region added on the right for clarity. Symbols show identified radial modes (red circles) and modelled radial and dipole modes (empty black circles and triangles, respectively).}
\label{fig:ech1}
\end{figure}

Hence, mode identification is the primary challenge for rapid rotators because it is a prerequisite for modelling the rotation rate. Regardless of rotation, the radial ($\ell=0$) pulsation modes of young intermediate-mass pulsators form a curved ridge up the centre of an \'echelle diagram of the observed pulsations (Fig.\,\ref{fig:ech1}), whilst dipole ($\ell=1$) modes usually form a comparatively straighter ridge separated from the radial modes by $\sim\Delta\nu/2$ \citep{beddingetal2020}. However, when a star is seen from an approximately equatorial inclination, the zonal modes (with azimuthal order $m=0$) are no longer visible, and only the rotationally-split sectoral modes ($|m| =\ell=1$) can be detected, complicating the mode identification further. At higher degrees, rotational splittings are the main source of additional ridges in \'echelle diagrams, such as those perceptible but unidentified in Fig.\,\ref{fig:ech1} that constitute the majority of the strong oscillation modes.

In this work, we analyzed the TESS light-curve of the multi-mode $\delta$~Scuti pulsator HD\,21434 (=TIC308243453), which is a member of FH. Only the radial modes could be identified. We modeled these using non-rotating models computed with MESA (r15140; \citealt{paxtonetal2011,paxtonetal2013,paxtonetal2015,paxtonetal2018,paxtonetal2019}) and pulsation calculations made with GYRE (v6.0.1; \citealt{townsend&teitler2013}). Our models are identical to those of SPYGLASS-II \citep{Kerr22}.
%Kerr et al CFN paper
We applied a solar-metallicity prior, [Fe/H] $= 0.0 \pm 0.1$\,dex, corresponding to metal mass fractions $Z_{\rm in} = 0.0130$ to 0.0160.

The \'echelle diagram (Fig.\,\ref{fig:ech1}) shows a very good fit to the radial modes, but the uncertain-yet-rapid equatorial rotation rate inferred from the projected value of $v \sin i = 93\pm4$\,km.s$^{-1}$ (this work) clouds the picture. In other words, the non-rotating models recover the mean density of the star asteroseismically, at 0.47\,$\rho_{\odot}$, but the mapping of stellar properties (mass, metallicity, age, and rotation) that gives this density is unconstrained because the equatorial rotation rate is unknown. We can, however, confirm that non-rotating models of the correct metallicity have pulsation frequencies consistent with the observed ones at ages of 20\,Myr. In the absence of rotation, the best-fitting models are the most metal-rich (least dense) in the grid, which suggests that rotating models whose density is lower due to centrifugal force could potentially be found within the same parameter range and at reasonable ages.

Although the current barrier to using rotating models is the identification of rotational splittings to constrain the rotation rate, we can make generic rotating models to investigate the influence of rotation. For instance, by computing a series models of a single mass and metallicity but different values of surface rotation, we found that the stellar density is about 10\,\% lower for a star with $v_{\rm eq}=150$\,km.s$^{-1}$ than for a non-rotating star. Since $\Delta\nu$ scales as $\sqrt{\rho}$ \citep{ulrich1986,kjeldsen&bedding1995}, this implies that the pulsation frequencies should be about 5\% lower in this rotating star. By scaling the frequencies of non-rotating models down by 5\%, we can therefore estimate a range of stellar ages within our non-rotating model grid that would reasonably reproduce the observed frequencies if the star were rotating. In doing so, we found that seven of the 50 top models have ages in the range $40\pm1$\,Myr, consistent with ages from Li depletion and the Tuc-Hor sequence in \citet{Kraus14} ($38.5\pm3.6$\,Myr), while the other 43 of the top 50 models have ages in the 20--25\,Myr range.

\subsubsection{Adopted Ages} \label{sec:ageadopted}

The various age sources presented through this work enable ages to be synthesized in a manner that makes use of the strengths of each. In this case, a combination of the isochronal and lithium depletion seems to produce the most coherent synthesis ages. The isochrones are capable of producing a robust formation sequence, while the lithium depletion ages are able to ensure that the absolute scaling is appropriate. Due to intrinsic features of the Austral association, including the presence of at least one marginally virialized cluster and its relatively old age range compared to CFN in SPYGLASS-II, the dynamical ages are likely not reliable for all subgroups. They are nonetheless useful for comparing with other methods. Similarly, the asteroseismic results will likely require additional rotational modelling to produce reliable results, so in this paper we exclude those results from our final age solutions. 

To compute combined ages, we begin by synthesizing a combined isochronal age out of the three models we use in this paper. Following the method described in detail in SPYGLASS-II, we fit linear relationships between the DSEP-Magnetic Ages and each of the PARSEC and BHAC15 ages using an orthogonal distance regression fitting routine. The combination of these fits produces a line in 3D isochronal age-space, which effectively averages over systematic relative biases between models. To produce an age, we must read off from this 3D line along the axis that best fits the lithium depletion age solutions, which we use to set our absolute scaling. The DSEP-magnetic models were designed in large part to produce agreement in age between lithium depletion and isochronal age solutions, and this case reflects that intention, with an age of $\sim$47 Myr produced for Tuc-Hor being roughly in agreement with range of possible ages for the region from \citet{Kraus14}, and much closer than the $\sim$20 Myr age solution from the BHAC15 models. We therefore read out from the line fit along the DSEP-magnetic axis, producing combined isochronal ages with a tether to the lithium depletion ages which ensures approximate agreement of their absolute ages. 

The resulting age solutions refine the age sequence in the region, giving Carina and Columba very similar ages at $\sim$26 Myr, and FH a somewhat older age at $\sim$32 Myr. Tuc-Hor is notably older, with an age of $\sim$46 Myr. The dynamical ages are generally not in dramatic disagreement with these results. The adopted ages and dynamical ages in Carina and Columba agree within uncertainties, while in the case of Columba the dynamical age is $\sim$1.2 Myr younger than the adopted age. This age gap can be interpreted as a dispersal delay caused by the presence of a significant gas mass soon after formation, and timescales $1<\tau<4$ Myr are consistent with both theoretical timescales \citep{Guszejnov22} and observational results in CFN from SPYGLASS-II. The results in FH and Tuc-Hor are less consistent. However, these results are unsurprising especially in FH, where inconsistent timescales of stellar dispersal are expected due to the plausibly virialized $\chi^{1}$ For cluster at the region's core. While the presence of a virialized core has not been established in Tuc-Hor due to the lack of a complete SPYGLASS survey there, the fact that the region has a notably older adopted age while maintaining a visibly dense core region suggests that even if it is not currently bound, it probably was in the past. The combination of past binding and a relatively old age would also provide opportunities for constituent subgroups to assemble hierarchically \citep{Guszejnov22}, potentially producing compact configurations in the region's past not caused by a compact state at formation. We therefore conclude that, while the dynamical ages do not perfectly agree with the adopted ages, they also do not provide significant tension. 

Carina is worth addressing separately, being the only region with a meaningfully different set of lithium EWs, although it does not show the same robust distinctions using other methods. While there are clear examples of stars in Carina with lithium measurements consistent with a younger age, Carina also has an isochronal age almost identical to Columba, and a secondary lithium sequence in line with the other Austral sub-associations, indistinguishable from the other sequences for $G_{BP}-G_{RP}<1.5$. Three Carina members have lithium measurements that suggest a younger age, but this is not a large enough sample to justify their use in place of those members on the older sequence. While our clustering results did indicate that Carina and Tuc-Hor members may be especially easy to mistake for one another, only one star in each is identified as a member of the other in the literature. While swapping these stars would help to solidify the younger sequence in Carina, this choice still does not completely dispel the validity of the older sequence. A possible solution is that there is a small young population overlapping in parameter space with Tuc-Hor and Carina that is younger, although too sparse to stand out against the denser populations in the area. This would explain the presence of the younger sequence and its apparent mixing with older populations in Tuc-Hor and Carina. It could also help to explain the isochronal and dynamical ages of Carina, since, if there were younger and older populations, our 26 Myr combined isochrone-based age for Carina could plausibly decompose into fits consistent with the $\sim$22 Myr younger lithium age sequence and the 32.9 Myr dynamical age. Investigating whether Carina contains an embedded younger population will require more complete spectroscopic coverage to more firmly establish the lithium sequence or sequences present, while refining RVs to improve membership vetting. While the age spread of the Carina sequence is not inconsistent with two differently-aged populations being present, it is also possible that improved membership vetting will resolve the contradictions in Carina's age. 

\subsection{Traceback} \label{sec:traceback}

\subsubsection{Connections with Other Nearby Associations} \label{sec:exttraceback}

Prior to performing a complete traceback on members of our combined Austral complex, it is important to assess the possibility of connections with other associations mentioned in the literature, especially those with previously claimed links to the populations we explore. Most notable among these is 32 Orionis, which was previously linked to Columba through the unsupervised machine learning clustering produced by \citet{LeeSong19}. One of the groups identified in that publication consisted of the whole of 32 Ori plus a subset of Columba members, hinting at a possible bridge between the populations. There are many other associations with similar ages to the Austral complex in the solar neighborhood and, while these have been more consistently separated from the populations we consider here in velocity space, it is worth considering them in case their orbits are convergent when traced back. For the purposes of this analysis, we considered Argus and $\beta$ Pic, in addition to 32 Ori. We also included Group 6 in this analysis due to it being centered on to the Sco-Cen-connected cluster Platais 8 (see Sec.~\ref{sec:g6}). SPYGLASS-I showed that $\beta$ Pic also has connections to Sco-Cen, so it and Group 6 will serve as two test cases for connections between Sco-Cen and the Austral complex. 

For this analysis, we used the same galpy numerical integration used for the dynamical ages \citep{Bovy15}, using inputs of RA, Dec, distance, proper motions, and RVs to ensure consistency between galpy's galactocentric plane and the heliocentric frame often used to describe these populations. \citet{Bell17} provided these mean values for 32 Orionis, while Argus and $\beta$ Pic are both notably closer to the Sun, making their velocity axes geometry-dominated. We therefore collected members for both Argus and $\beta$ Pic and computed their trajectories before averaging. We used the catalogs from \citet{Malo13} and \citet{Shkolnik17}, removing any sources with evidence of binarity, either through a resolved companion or RUWE$>$1.2. Since we already have members with velocities for Group 6, we followed the same approach for them, tracing the members forward and averaging at each step. We then assessed connections to the Austral complex by performing the same traceback on its well-established components, and computing the distances between cores in the Austral complex and those in candidate external populations.

None of Argus, $\beta$ Pic, or 32 Orionis had close approaches to the Austral complex around the time its constituent subgroups were forming. Nearly all of these groups only had close approaches to the Austral complex within the last 10 Myr, with the only exception being the pair of 32 Orionis and Tuc-Hor, whose closest approach was 37 Myr ago, albeit at a distance of over 90 pc. During the most recent burst of star formation, which formed Carina, Columba, and FH, all mutual distances were over 100 pc. We therefore conclude there are unlikely to be any direct connections between the Austral complex and these associations. The clear separation between 32 Orionis and Columba, in particular, is notable due to their proposed merger in \citet{LeeSong19}. This connection may have been produced by an imperfect training set, given there are a few proposed Columba members used in that clustering analysis that partially encircle 32 Ori in spatial coordinates.

Connections between the Austral complex and Group 6 cannot be dismissed as easily. Mutual distances 30--40 Myr ago between Group 6, Carina, and Columba were less than 50 pc at times which, while not implying overlap, does make this separation comparable to those we show between Tuc-Hor and the other Austral sub-associations (see below). Due to the connection between Group 6 and Platais 8, a cluster which SPYGLASS-I grouped in with the incredibly complicated Sco-Cen Association, further analysis on this possible connection is beyond the scope of this paper. However, this connection should be considered in future investigations of the formation of Sco-Cen, especially in the dynamically distinct region that contains Platais 8, which SPYGLASS-I refers to as the IC 2602 branch. 

\subsubsection{Star Formation History}

\begin{figure*}
\centering
\includegraphics[width=15cm]{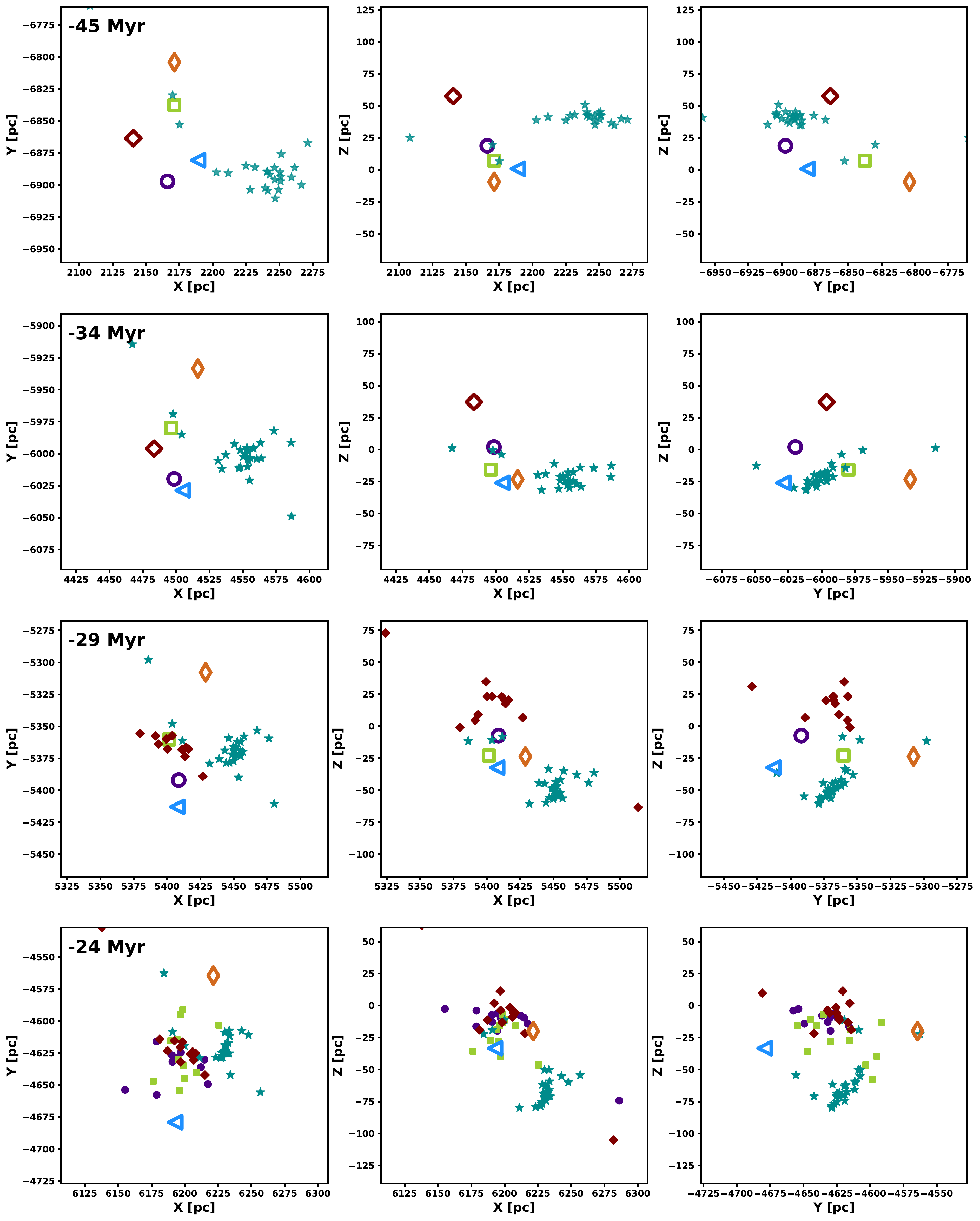}\hfill
\caption{Locations of stars in the Austral complex over the course of its formation. The marker shapes and colors match those in Figure \ref{fig:australspatial}. Before stars form, we provide only the mean position of stars at that time step using large open markers, roughly indicating the motions of progenitor gas. After formation, we show individual stars used in traceback. We change the transparency of the symbols from invisible to solid linearly over the age uncertainty interval of the subgroup, such that stars within uncertainties of their adopted age are shown as transparent. The first two time steps ($-$45 Myr and $-$34 Myr) show the configuration at the beginning and end of the period where only Tuc-Hor was present, with FH forming soon after that second time step. The third time step ($-$29 Myr) shows the region after the formation of FH, while the final time step ($-$24 Myr) shows the region after the formation of Columba and Carina, which together form the last period of star formation. Groups 5 and 6 are always shown as single open icons, as they have weaker connections to the Austral complex and do not have age solutions. An interactive version is available in the online-only version, which includes complete traceback to the present day. \edit1{The interactive version can be manipulated in 3D using zoom, rotation, and reset, and it can be manipulated in time using the slider.}}
\label{fig:traceback}
\end{figure*}

By combining our ages for the Austral complex with the motions of the constituent stars, we can reconstruct the patterns of formation in the region. The complete traceback result is shown in Figure \ref{fig:traceback}, providing an overview of the motions of stars alongside their times of formation at select times. The full traceback sequence to the present day is available in the online-only version of that figure. 

Our results show the onset of star formation $\sim$46 Myr ago in Tuc-Hor, 14 Myr before the next group of stars formed (Fig \ref{fig:traceback}, row 1). During this period, none of the traced back locations of other subgroups were especially nearby, with Tuc-Hor separated from Carina and FH by over 100 pc, and Columba by nearly 90 pc. Group 5 was the most distant subgroup to Tuc-Hor at this stage, and did not get much closer later in the Austral Complex's formation, further disputing its connection to the Austral Complex. 

Approximately 32 Myr ago, or between rows 2 and 3 of Figure \ref{fig:traceback}, FH became the second subgroup in the Austral complex to form. At this stage, FH, Carina, and Columba had not significantly reduced their distance from Tuc-Hor. The formation of FH therefore took place over 80 pc away from Tuc-Hor, suggesting a relatively weak connection between the two. While Tuc-Hor remained well-separated from these other populations, the centers of Carina and Columba soon after converged on FH (row 4 of Fig. \ref{fig:traceback}), both forming just as the three populations began to overlap in spatial coordinates, around 26 Myr ago. This convergence suggests that the formation of Carina and Columba was intertwined with the formation of Fornax-Horologium, likely occupying the same molecular cloud, while Tuc-Hor appears to have formed mostly independently of the others. This result is similar to what was seen in SPYGLASS-II, where two distinctly different nodes of star formation were also present, like the emerging distinction between a ``Tuc-Hor node'' and an ``FCC node'' (Fornax-Horologium, Columba, Carina) in this work.  

\section{Stellar Populations} \label{sec:stellarpops}

\subsection{Final Sample} \label{sec:fhpops}

Using similar methods to those used in Section 5.3 of SPYGLASS-II, we can estimate the total mass and stellar population of the Fornax-Horologium Association, correcting for field contamination, unresolved binaries, and stars below the minimum mass considered in our young star identification. The conclusions will depend on the definition of FH used, and there are two slightly different definitions presented in this publication: the population recovered directly from our search for young populations, like what is shown for CFN in SPYGLASS-II, and the reclustered population, which introduces neighboring populations and reassigns stars to the most likely parent population. While the former is a reliable choice in more distant or isolated environments, the closeness of FH to the Sun means that the near edge of the cluster, which is defined by a drop in density, will be set by the dispersal of the 2D velocity vectors as a result of spherical geometry, not a genuine density drop. This is likely why Columba did not appear in SPYGLASS-I, despite our combined Austral population showing that the densities throughout much of FH and Columba are similar. We therefore conclude that the reclustered definition of FH is likely more accurate, as it sets the boundary between Austral sub-associations based on proximity to the nearest population, not continued elevated density in an environment where other overdense regions are present. 

We next estimated the contamination fraction in this population. While some of the reclustered FH members were initially identified as part of other Austral sub-associations, for which we have little information on contamination, 311 of 329 stars in this sample are also in the separate initial sample of 329 FH candidate members. This similarity between the populations implies that that initial FH population can be used to estimate the contamination in the reclustered sample. We therefore estimate the contamination of the full set of 654 space-velocity neighbors to the DR3-updated SPYGLASS-I sample, and the subset of 329 consistent with photometric youth. 

Stars between the 0.25 and 0.5 M$_{\odot}$ isomass curves have a clean separation in the color-magnitude diagram, so contamination can be reliably estimated by simply counting the number of stars between the two isomass curves that pass the photometric youth cut in Section \ref{sec:photocut}, versus those that do not, assuming that there are no mass-dependent anomalies to the initial mass function. We find that 76 of 189 stars in this mass range are photometrically young, suggesting that 40$\pm$7\% of stars in the photometrically unrestricted initial sample are Fornax-Horologium members, with uncertainties from binomial statistics. This corresponds to 263$\pm$46 expected members, compared to 654 in the full population and 329 stars in the photometrically restricted candidate population. Since the reclustered cluster definition is the same size, we estimate a total population of 253$\pm$44 members. 

We then assembled our population of 263 genuine members out of two components: the 208 stars we identified as photometrically young, and a subset of stars above the pre-main sequence turn-on with ambiguous youth containing the 55 members not found in the pre-main sequence population. This division implies that 55 of 121 (45\%) stars above the PMSTO are Fornax-Horologium members. We can therefore numerically remove contamination by applying a corrective factor of 0.45 to stars with ambiguous youth, which can be multiplied by both the implied mass contained in those stars and their populations. When 30 clear white dwarfs in the initial definition of FH are counted as high-mass field stars, a population of 55 FH members in this ambiguous region of the CMD implies a 36\% membership rate, quite consistent with the  contamination estimate among lower-mass sources. 

Next, we corrected the sample for binary populations. Many binaries are unresolved and therefore not detectable through Gaia, which consequently hides mass associated with the companion. The demographics of binary companions are becoming well-understood through recent works \citep[e.g.,][]{Raghavan10,Duchene13,Sullivan21}, so we can correct for unresolved binaries by restricting our sample to stars that our binary identification script identifies as either single or a primary, and reintroducing the expected mass and population of binaries expected from that population. Multiplicity and mass ratio are both properties that have been considered in recent publications, and combining this information with the estimated mass of the primary can provide a estimate for the averaged complete mass of the system. For multiplicity rates, we interpolated the results provided by \citet{Sullivan21} as a function of mass to estimate multiplicity for each individual star. We then calculated the mean mass ratio using the \citet{Sullivan21} power law indices for companion mass ratio for a star of a given mass, again interpolating between the masses provided. Multiplying the mean mass ratio, mean multiplicity, and primary mass together produced a mean missing mass, which can be added to the mass of the primary to get an expected system mass. Similarly, the mean multiplicity can be added to the population of primaries to estimate the number of stars in the population, including binary companions. After the correction to exclude field populations on the upper main sequence, this process added 19 M$_\odot$ to the population in FH. 

Finally, we adjusted our mass estimate for the presence of lower-mass objects, which have a mass less than the 0.09 M$_{\odot}$ isochronal mass limit in young star detection. SPGYLASS-II used the \citet{Chabrier05} IMF to estimate the missing stellar mass below this 0.09 M$_{\odot}$ mass cutoff, and found that 2\% of stellar mass is below this limit. We did not include inferred populations of lower-mass stars in our membership count, but note that the mass of these objects is important to understand when making estimations of binding in this association.

After making these adjustments, we conclude that Fornax-Horologium contains 273$\pm$48 stars across 198$\pm$35 stellar systems with M $>$ 0.09 M$_{\odot}$, totalling approximately 106$\pm$19 M$_{\odot}$. Our uncertainties are based on the fractional value in the input prior which conditions much of these estimates. Other sources of uncertainty, such as systematic uncertainties in mass estimation and the reintroduction of binaries especially at young ages \citep[e.g.,][]{Kraus11,Feiden16}, are much more difficult to quantify. A more complete overview of the sources of uncertainties in these estimations is provided in SPYGLASS-II. The choice of how FH is defined is also important here, especially due to ambiguous boundaries with neighboring Austral sub-associations. While our two different definitions of FH yield the same sizes, a definition that separates parts of the extended halo from the $\chi^1$ For cluster could result a very different population. 

Our catalog of credible candidates includes 329 objects with youth-consistent photometry, 250 of which also passed our RV cut. The number of stars that passed the photometric and radial velocity cuts is only slightly smaller than the association's expected total population, so we expect it to closely reflect the association's total membership. However, the presence of high-velocity binaries does introduce the possibility of genuine members being excluded from the sample and non-members being included. The \citet{Galli21} catalog, which provided the most complete view of FH prior to this publication, contains 164 objects, including all but four of the candidates found in the next most complete catalog \citep[published by][]{Zuckerman19}. Our catalog contains 144 out of 164 (87\%) suspected members from \citet{Galli21}, and all but two of the excluded stars were removed by our photometric youth cut. The remaining 185 candidate FH members in our catalog are all newly-identified, and 120 of these passed our RV cut. We find the \citet{Galli21} catalog to be nearly complete towards the $\chi^{1}$ For cluster centre, but our catalog provides a significant expansion on the edges, made possible by the SPYGLASS program's sensitivity to extended structures. This low-density envelope is the region of Fornax-Horologium that is least well-understood, and it begins to blend into the Columba association at its northern frontier, where an extended tail reaches into Columba while containing a small number of proposed Columba members. This result further motivates the assessment of the relations between these two populations. 

\subsection{Virial State of Fornax-Horologium} \label{sec:fhvir}

The interpretation of Fornax-Horologium's dynamical ages will depend on the group's virial state. A bound cluster will not converge as its members are traced back in time, because the motions of those members are dominated by random motions in the cluster produced by gravitational interactions, rather than simple dispersal. We therefore investigated the virial state of Fornax-Horologium using the methods previously employed in SPYGLASS-II for the EE Draconis cluster, following the methods of \citet{PortegiesZwart10} and \citet{Kuhn19}. Unbound associations satisfy the inequality $\sigma_{1 \rm D} > \sqrt{2} \sigma_{\rm virial}$, where $\sigma_{1 \rm D}$ is a characteristic 1-D velocity dispersion, and $\sigma_{\rm virial}$ is the virial velocity. 

Following SPYGLASS-II, we calculated both values for the entirety of Fornax-Horologium. For $\sigma_{1 \rm D}$, we computed the value as the square root of the mean variance across $\Delta$v$_{T, \rm RA}$ and $\Delta$v$_{T, \rm Dec}$, clipping at 2-sigma to avoid outliers. We also excluded binaries to avoid any contributions from internal motions. The result was $\sigma_{1 \rm D} = 0.32$ km s$^{-1}$. We then computed $\sigma_{\rm virial}$. Using a density profile parameter $\eta = 10$, which assumes a Plummer profile, we computed $\sigma_{\rm virial} = 0.07$ km s$^{-1}$, while assumptions of a looser density profile $\eta = 5$ provide a value of $\sqrt{2} \sigma_{\rm virial} = 0.1$ km s$^{-1}$. None of these values is close to suggesting a bound state for Fornax-Horologium, given a value of $\sigma_{1 \rm D} = 0.32$ km s$^{-1}$. 

While the above result confirms that Fornax-Horologium is in large part an unbound group, this does not suggest that no section of the association is bound. In fact, visual inspection reveals a dense core in spatial coordinates that is reflected in the velocity distribution, suggesting that there may still be a virialized core in the center of the association. This core is centered on $\chi^{1}$ Fornacis, and extends about 5 pc from that star, and stars within that 5 pc radius have a $\sigma_{1 \rm D} = 0.09 $ km s$^{-1}$. The corresponding $\sqrt{2} \sigma_{\rm virial}$ values using only the mass in that region are $\sqrt{2} \sigma_{\rm virial} = 0.06$ km s$^{-1}$ for $\eta = 10$, and $\sqrt{2} \sigma_{\rm virial} = 0.08$ km s$^{-1}$ for $\eta = 5$. This puts the state of the cluster core in a plausibly virialized regime. Furthermore, only 9.4 M$_\odot$ reside within that radius out of the 106 M$_\odot$ in the association, suggesting that the core currently observed is a heavily stripped remnant of what the cluster used to be, with many stars having been ejected from that core since formation. This could explain why Fornax-Horologium is seen here as only marginally virialized, since the core may have recently been much more massive, further supporting binding. 

We therefore conclude that, while the vast majority of FH is unbound and divergent, there is a core region within FH with strong evidence for current or at least past virialization. We therefore conclude that Fornax-Horologium is a dissolving cluster in which a small virialized core still exists, but most members have been ejected from the core since formation. This distinction between the core and outer regions also motivates a change in the naming convention for the region. Where $\chi^{1}$ For and Alessi 13 remain appropriate designations for the association's marginally bound core, it is useful to have a separate name that includes the entire extended halo, which is what we refer to as Fornax-Horologium, following SPYGLASS-I.

In a mostly unbound association with a potentially bound central core, dynamical traceback to a tighter past configuration is not expected to be a reliable indicator of age, since the stars traced back will have an ejection time spread throughout Fornax-Horologium's history. However, some stars will have been ejected from the association early in formation, and there should therefore be a limited population capable of providing a lower limit of the association's age. 

\section{Discussion} \label{sec:discussion}

The sequence of formation seen in the Austral complex is not unlike what SPYGLASS-II observed in CFN, in which star formation events took place in distinctly different nodes. In the Austral complex, two such nodes emerged, both forming stars co-spatially, with one node containing just the Tuc-Hor association (Tuc-Hor Node), and the other producing FH, Columba, and Carina (FCC Node). In CFN, this dual-node structure was explained as the possible consequence of a fragmenting filament, in which material along the filament established two collection points along its length, eventually forming stars. The overall separation between these nodes is larger than the $\sim$30 pc seen in CFN, with separations between 50 and 70 pc when Carina and Columba formed, and a more significant $\sim$85 pc gap to FH when it formed. While this is certainly a less direct connection than was seen between the formation nodes of CFN, it is still within the range of distances seen in the known dense filaments around the solar system \citep[e.g.,][]{Goodman14, Zucker15}, so this does not discount the filamentary formation model. 

Both nodes show an age distribution consistent with continuous formation. The Tuc-Hor node does not show physical substructure of any form, suggesting only one generation exists, while the FCC node had a relatively short 6 Myr star formation period, shorter than the span of ages in either CFN node in SPYGLASS-II. This lifespan is well within the range expected for individual star forming events \citep[][]{Guszejnov22}, suggesting that any gaps seen between ages could easily be spanned by the intrinsic age spreads within individual subregions of a molecular cloud. Careful work on computing age spreads will be useful to verify this, although this requires precise knowledge of the contamination from field stars, interlopers from other subgroups, and binaries, while also requiring a very accurate view of the uncertainties involved in the astrometry and photometry, making it beyond the scope of this publication. While star formation in individual nodes appears to have been continuous, we find a very significant 14-Myr age gap between star formation in Tuc-Hor and FH, wide enough that direct connection between the star formation events is unlikely. This is perhaps to be expected, even if they formed in the same filament, as the two nodes formed stars far enough apart that they likely did not experience the same conditions throughout their evolution.

While this filament-based view of star formation in the Austral complex explains the region well in isolation, it is complicated by the proximity of Group 6, which is mostly comprised of stars in and around Platais 8 in Sco-Cen. During the formation of the other Austral populations, the location of Platais 8 remained relatively central within the Austral Complex, making its exclusion difficult to justify at the same time the Tuc-Hor and FCC nodes are grouped together. Our work is not the first to note the dynamical similarities between Platais 8 and components of the Austral Complex.
\citet{Gagne21} used kinematics and color-magnitude diagrams to suggest that Carina and Columba are connected to with Platais 8 instead Tuc-Hor, and further proposed that Tuc-Hor is instead connected to IC 2602. Both Platais 8 and IC 2602 are on the IC 2602 branch of Sco-Cen in SPYGLASS-I, suggesting that the IC 2602 branch and Austral Complex may have a closely intertwined history. Furthermore, a dynamically similar population (MELANGE-4) filling most of the space between LCC in Sco-Cen and Carina was identified by \citet{Wood22}, providing a possible link between the Austral Complex and the rest of Sco-Cen. While a complete assessment of the relationship between Sco-Cen and the Austral sub-associations is beyond the scope of this publication, the complicated web of proposed connections emerging suggests that the formation of the two regions may be directly connected, perhaps making a single filament explanation overly simplistic. This fact demonstrates that, while views of isolated star formation events can provide useful insights into star formation processes, viewing localized groups or even entire complexes in isolation may miss connections at larger scales.

Star formation is a complicated process that often resists rigid divisions between events. However, this publication and SPYGLASS-II are beginning to show that star formation nodes may represent a strong and coherent unit of star formation, providing discrete positions and times where star formation took place. Most of the nodes described so far have contained multiple associations or constituent populations without much ambiguity in assigning populations to parent nodes, making nodes perhaps the most coherent building block of larger-scale star formation events. Future advancements in large-scale star formation may therefore benefit from this node-based breakdown of the process. By using dynamical traceback to establish the positions and active periods of star formation from multiple nodes across many of the traditionally defined star formation complexes, we may be able to establish connections that are not clear from views of space-velocity distributions without traceback. Much broader coverage of nearby associations will nonetheless be necessary to produce a sample sufficient to assemble some of the larger local star formation events from smaller nodes.

\section{Conclusion} \label{sec:conclusion}

We have greatly expanded the known populations around the $\chi^1$ Fornacis cluster, establishing a much broader association around the central cluster, which we refer to as the Fornax-Horologium Association (FH). We identified 329 candidate members, which we merged with literature samples of Tuc-Hor, Columba, and Carina to produce an aggregate sample for the entire Austral complex containing 811 stars. In doing so, we have provided the first view of these populations not as disparate groups, but as a continuous network. The key findings of the subsequent analysis are as follows: 
\begin{enumerate}
    \item Re-clustering within the Austral Complex recovered the four known populations of FH, Tuc-Hor, Columba, and Carina, as well as two additional populations: Group 6, which is centered on the Sco-Cen-linked Platais 8 cluster, and Group 5, which is a more tenuous group that requires verification. 
    \item We find that non-Gaia RV sources for the members of the proposed Smethells 165 moving group show velocities very consistent with other Tuc-Hor members, suggesting that this group is spurious, and a likely consequence of artefacts in Gaia RV data.
    \item Star formation in the Austral complex divides into two distinct nodes of co-spatial formation, with constituent ages consistent with continuous formation: the Tuc-Hor node around its namesake association, and the FCC node, which contains FH, Columba, and Carina.
    \item Group 6 remained close to the Austral complex throughout its formation, despite its connections to the Sco-Cen complex, suggesting that the Austral complex may be only a small portion of a much larger network of intertwined nearby associations.
\end{enumerate}

Our results in the Austral complex provide a continuation of previous work in SPYGLASS-II, building a view of star formation with nodes -- continuous and cospatial hubs of star formation -- as the most distinctive discrete components. Traceback is essential to identifying these nodes, since their constituent associations are not always trivial to connect using present-day space-velocity distributions alone. The potential connection between the Austral Complex and Sco-Cen enforces the value of expanding these studies in a manner that can establish more of these nodes, providing a promising avenue to establish star formation patterns spanning multiple associations. 

\begin{acknowledgments}

Funding for RMPK and ACR was provided by the Heising-Simons Foundation. RMPK acknowledges the use of computational  resources  at  the  Texas  Advanced Computing Center (TACC) at the University of Texas at Austin, which was used for the more computationally intensive operations in this project. RMPK acknowledges the help of Ben Tofflemire, whose expertise was instrumental in enabling our RV measurements. TRB and SJM gratefully acknowledge support from the Australian Research Council through Discovery Project DP210103119, Future Fellowship FT210100485 and Laureate Fellowship FL220100117.

\end{acknowledgments}

\vspace{5mm}
\facilities{Gaia, Las Cumbres Observatory: NRES Spectrograph on 1m Telescope at Sutherland (CPT), Wise (TLV), and Cerro Tololo (LSC)}

\software{astropy \citep{Astropy13},  
          {\tt saphires} \citep{Tofflemire19}, echelle \citep{hey&ball2020}}
          
\appendix
\section{Binaries} \label{app:bin}

This appendix provides a catalog of likely binaries in the Austral complex, similar to that shown in SPYGLASS-II. We include both stars in our new SPYGLASS-based sample for FH and the stars added through a search of the literature. We identify objects as binaries when they have a companion within 10000 AU in the plane of the sky at the star's distance, provided that they have parallaxes within 20\% and proper motions within 5 km s$^{-1}$. In cases where an object was found within the 10000 AU search radius, it is only recorded if the star being searched is brighter than its companions. This avoids duplicates and ensures that the search is focused on the primary, avoiding chains of binary affiliation from developing. We identify 206 stars in a binary or multiple system, of which 53 are not in our candidate member catalog. This relatively large number may suggest that, especially outside of our SPYGLASS-set sample in FH, the other subregions of this complex may need their coverage expanded. High internal velocities in some binary systems will, however, occasionally make memberships of such systems difficult to kinematically verify, even with deeper coverage. 

\input{binaries}

\bibliography{rkerr_refs,sjm_refs}{}
\bibliographystyle{aasjournal}

\end{document}

%% file: AUSTRAL_SPECTRA.tex
\begin{deluxetable*}{cccccccccccc}
\tablecolumns{12}
\tablewidth{0pt}
\tabletypesize{\scriptsize}
\tablecaption{Spectroscopic properties of members in Fornax-Horologium and the connected Tuc-Hor, Columba, and Carina Associations, including both literature values and new observations. The source for each value is included. High-quality literature values supercede our observations in cases of poor RV results or low-signal line measurements. The complete version of this table is available in the online-only version, which contains 628 entries}
\label{tab:spectres}
\tablehead{
\colhead{Gaia ID} &
\colhead{RA} &
\colhead{Dec} &
\multicolumn{3}{c}{RV (km s$^{-1}$)} &
\multicolumn{3}{c}{EW$_{Li}$ (\AA)} &
\multicolumn{3}{c}{EW$_{H\beta}$ (\AA)} \\
\colhead{} &
\colhead{(deg)} &
\colhead{(deg)} &
\colhead{val} &
\colhead{err} &
\colhead{src\tablenotemark{a}} &
\colhead{val} &
\colhead{err} &
\colhead{src\tablenotemark{a}} &
\colhead{val} &
\colhead{err} &
\colhead{src\tablenotemark{a}} \\
}
\startdata
 4855470634187962496 &   57.6708 & -39.7451 &  17.31 &   10.52 &         DR3 &        &         &       &        &         &      \\
 4855535127416837888 &   57.1700 & -39.4083 &  17.41 &    2.63 &         DR3 &        &         &       &        &         &      \\
 4870445020485171712 &   63.8883 & -34.1556 &  23.29 &    1.11 &         DR3 &        &         &       &        &         &      \\
 4846243493951342336 &   49.8962 & -46.6763 &  60.00 &    0.98 &         DR3 &        &         &       &        &         &      \\
 4860290961883535360 &   54.5367 & -35.5719 &  16.96 &    4.76 &         DR3 &        &         &       &        &         &      \\
 5058474290658453888 &   48.3290 & -29.8887 &  46.40 &    0.17 &         DR3 &        &         &       &        &         &      \\
 5094142482222743936 &   60.4671 & -20.4750 &   5.30 &    0.20 &          K7 &        &         &       &        &         &      \\
 4855882229493582080 &   57.8234 & -38.2101 &  33.97 &    0.18 &         DR3 &        &         &       &        &         &      \\
 4864858814221301504 &   68.0339 & -38.2958 &  22.94 &    0.09 &         LSC &  0.209 &   0.020 &   LSC &   0.00 &    0.00 &  LSC \\
 5056460157154931584 &   52.4042 & -30.7308 &  10.85 &    1.14 &         DR3 &  0.026 &   0.055 &   LSC &   0.00 &    0.00 &  LSC \\
 4858398152615945600 &   57.6102 & -35.7675 &  14.85 &    5.09 &         DR3 &        &         &       &        &         &      \\
 4862215863146322432 &   57.8252 & -33.7499 &  30.66 &    0.05 &         CPT &  0.021 &   0.018 &   CPT &   0.00 &    0.00 &  CPT \\
\enddata
\tablenotetext{a}{The source for the measurement, either the LCO node used for an original observation (CPT, LSC, or TLV), or an external source. The abbreviations are: Gaia DR3 (DR3),  \citet{Grenier99} (G99), \citet{Gizis02} (Gi), \citet{Bobylev06} (B6), \citet{Gontcharov06} (G6), \citet{Riaz06} (R), \citet{Torres06} (T6), \citet{White07} (W), \citet{Kharchenko07} (K7), \citet{Mentuch08} (Me), \citet{daSilva09} (D), \citet{Casagrande11} (C11), \citet{Kiss11} (Ki) \citet{deBruijne12} (dB12), \citet{Anderson12} (A12), \citet{LopezMarti13} (L13), \citet{Rodriguez13} (Ro), \citet{Moor13} (Mo13), \citet{Malo13} (M) \citet{Elliot14} (E), \citet{Malo14} (M14), \citet{Kraus14} (K), \citet{Burgasser15} (B15), \citet{Desidera15} (D15), \citet{Faherty16} (F16), \citet{RAVEKunder17} (RAVE5), \citet{Shkolnik17} (S17), \citet{Fouque18} (F18), \citet{Gagne18BXI} (11), \citet{Gagne18BXII} (12), \citet{Gagne18BXIII} (13), \citet{Jeffers18} (J18), \citet{Kounkel18} (K18), \citet{Soubiran18} (S18), \citet{Bowler19} (B), \citet{Flaherty19} (F19), \citet{Schneider19} (S), \citet{Sperauskas19} (Sp19), \citet{Xiang19} (X19), \citet{Steinmetz20} (RAVE6)}
\vspace*{0.1in}
\end{deluxetable*}

%% file: AUSTRAL_MEM.tex
\begin{deluxetable*}{cccccccccccccccc}
\tablecolumns{16}
\tablewidth{0pt}
\tabletypesize{\scriptsize}
\tablecaption{Stars identified as credible candidate members of Fornax-Horologium and adjoining Austral populations, shown alongside Gaia sky positions, distances, photometric data, and clustering proximity ($D$), which indicates proximity to other group members. We also include each star's parent subgroup, estimated mass, and various quality flags, covering photometric membership, velocity membership, binarity, and overall quality, which are used to produce the higher-confidence samples used in later analyses (see Section \ref{sec:sss}). The complete version of this table is available in the online-only version, which contains all 811 credible Austral complex members.}
\label{tab:australproperties}
\tablehead{
\colhead{Gaia ID} &
\colhead{IG\tablenotemark{a}} &
\colhead{SRC\tablenotemark{b}} &
\colhead{REG\tablenotemark{c}} &
\colhead{RA} &
\colhead{Dec} &
\colhead{$\pi$} &
\colhead{$m_G$} &
\colhead{$G_{BP}-G_{RP}$} &
\colhead{M} &
\colhead{$D$} &
\colhead{$A$\tablenotemark{d}} &
\colhead{$P$\tablenotemark{e}} &
\colhead{$PY$\tablenotemark{f}} &
\colhead{$V$\tablenotemark{g}} &
\colhead{$F$\tablenotemark{h}} \\
\colhead{} &
\colhead{} &
\colhead{} &
\colhead{} &
\colhead{(deg)} &
\colhead{(deg)} &
\colhead{(mas)} &
\colhead{} &
\colhead{} &
\colhead{(M$_{\odot}$)} &
\colhead{} &
\colhead{} &
\colhead{} &
\colhead{} &
\colhead{} &
\colhead{}
}
\startdata
 4882777967536607616 &   FH &    F &       4 &   62.5444 & -32.9700 &  10.02 &  0.72 &      7.00 &  1.14 &       0.02 &                 1 &                 1 &         0 &       1 &    0 \\
 5081320011979189120 &   FH &    F &       4 &   54.8587 & -27.5004 &  16.37 &  3.03 &      7.00 &  0.26 &       0.06 &                 1 &                 1 &         1 &       0 &    0 \\
 5081415806929560448 &   FH &    F &       4 &   55.0451 & -27.3674 &  16.74 &  3.14 &      7.07 &  0.23 &       0.07 &                 1 &                 1 &         1 &       0 &    1 \\
 5081415806929560704 &   FH &    F &       4 &   55.0455 & -27.3687 &  16.64 &  3.11 &      7.08 &  0.24 &       0.08 &                 1 &                 1 &         1 &       0 &    1 \\
 4863772118776214400 &   FH &    F &       4 &   56.1203 & -30.9173 &   9.10 &  0.62 &      7.11 &  1.30 &       0.05 &                 1 &                 1 &         0 &      -1 &    9 \\
 4885553920862382720 &   FH &    F &       4 &   63.9524 & -29.5398 &  12.50 &  1.34 &      7.21 &  0.72 &       0.02 &                 1 &                 1 &         0 &      -1 &    0 \\
 4868955766345648768 &   FH &    F &       4 &   66.0504 & -37.0299 &  14.04 &  2.63 &      7.27 &  0.47 &       0.04 &                 1 &                 1 &         1 &      -1 &    0 \\
 5085616967843502976 &   FH &    F &       4 &   55.1141 & -23.7618 &  18.33 &  3.60 &      7.34 &  0.14 &       0.09 &                 1 &                 1 &         1 &       0 &    0 \\
 4855535127416838016 &   FH &    F &       4 &   57.1708 & -39.4095 &  13.28 &  1.96 &      7.36 &  0.68 &       0.07 &                 1 &                 1 &         0 &       1 &    1 \\
 4855470634187962496 &   FH &    F &       4 &   57.6708 & -39.7451 &  14.71 &  2.65 &      7.38 &  0.44 &       0.07 &                 1 &                 1 &         1 &       1 &    0 \\
 4855535127416837888 &   FH &    F &       4 &   57.1700 & -39.4083 &  14.45 &  2.46 &      7.38 &  0.52 &       0.09 &                 1 &                 1 &         1 &       1 &    1 \\
 4870445020485171712 &   FH &    F &       4 &   63.8883 & -34.1556 &   9.02 &  0.54 &      7.39 &  1.36 &       0.06 &                 1 &                 1 &         0 &       1 &    1 \\
\enddata
\tablenotetext{a}{The literature group the star was initially assigned to. Stars assigned to FH constitute the candidate member sample for Fornax-Horologium produced by this publication}
\tablenotetext{b}{The source the star was drawn from. The abbreviations are: the FH sample from this work (F), \citet{Torres08} (T), \citet{Malo13} (M), \citet{Kraus14} (K), \citet{Schneider19} (S), \citet{Gagne18BXI} (11), \citet{Gagne18BXII} (12), \citet{Gagne18BXIII} (13), and \citet{Booth21} (B21)}
\tablenotetext{c}{The ID of the HDBSCAN subgroup the star is assigned to. Group 1 corresponds to Columba, 2 is Carina, 3 is Tuc-Hor, and 4 is FH.}
\tablenotetext{d}{The boolean solution to the Astrometric quality cut from SPYGLASS-I, which is based on the unit weight error. 1 passes, 0 fails. }
\tablenotetext{e}{The boolean solution to the Photometric quality cut from SPYGLASS-I, which is based on the BP/RP flux excess factor. 1 passes, 0 fails.}
\tablenotetext{f}{A flag for the photometric youth determination of the star. A value of 1 passes, 0 is inconclusive. Stars that fail this cut are not included in this table.}
\tablenotetext{g}{A flag to represent the results of our velocity membership cuts. 1 passes, -1 fails, and 0 has no RV.}
\tablenotetext{h}{A flag for other notable features. 1 indicates that the star has a resolved companion within 10000 AU in the plane of the sky, 2 indicates a bad broadening function solution, 4 indicates a bimodal line profile likely indicative of spectroscopic binarity, 8 indicates an RUWE$>$1.2, indicating likely unresolved binarity. The flags are added in cases where multiple are true; for example, flag 6 indicates both flag 2 and 4. }
\vspace*{0.1in}
\end{deluxetable*}

%% file: SG_PROPS.tex
\begin{deluxetable*}{cccccccccccc}
\tablecolumns{12}
\tablewidth{0pt}
\tabletypesize{\scriptsize}
\tablecaption{The results of re-clustering in the Austral complex, providing basic properties across all proposed populations in the Austral complex, including Groups 5 and 6, which still need their existence and connection to the rest of the Austral complex verified. We provide the parent node (FCCN -- FH-Columba-Carina Node or THAN -- Tuc-Hor Association Node) and number of members, alongside the mean distance, mean position in 3D XYZ galactic coordinates, velocities in UVW space, and the age, if available. }
\label{tab:sgprops}
\tablehead{
\colhead{SG} &
\colhead{Node} &
\colhead{Name} &
\colhead{N$_{stars}$} &
 \colhead{d} &
 \colhead{X} &
 \colhead{Y} &
 \colhead{Z} &
 \colhead{U} &
 \colhead{V} &
 \colhead{W} &
\colhead{Age} \\
\colhead{} &
\colhead{} &
\colhead{(deg)} &
\colhead{(deg)} &
\colhead{(pc)} &
\colhead{(pc)} &
\colhead{(pc)} &
\colhead{(pc)} &
\colhead{(km s$^{-1}$)} &
\colhead{(km s$^{-1}$)} &
\colhead{(km s$^{-1}$)} &
\colhead{(Myr)}
}
\startdata
1 & FCCN & Columba &  124 &   71.0 & -34.0 &  -44.0 & -31.0 & -12.80 & -21.52 & -5.27 &        26.4 $\pm$ 1.7 \\
2 & FCCN & Carina  &  101 &   76.0 &  14.0 &  -70.0 & -19.0 & -10.35 & -22.35 & -5.52 &        26.4 $\pm$ 1.7 \\
3 & THAN & Tuc-Hor &  220 &   46.0 &  12.0 &  -24.0 & -35.0 &  -9.49 & -21.10 & -1.01 &        46.3 $\pm$ 2.3 \\
4 & FCCN & FH      &  329 &  108.0 & -33.0 &  -49.0 & -87.0 & -12.57 & -21.89 & -4.55 &        32.4 $\pm$ 1.3 \\
5 & N/A & -- &   22 &   42.0 & -30.0 &   20.0 &  -1.0 & -11.14 & -22.74 & -5.78 &     -- \\
6 &  N/A & -- &   15 &  129.0 &   7.0 & -122.0 & -15.0 & -11.34 & -22.93 & -4.00 &            -- \\
\enddata
%\tablenotetext{a}{Has a close approach with Tuc-Hor Node around the time of Tuc-Hor's formation, but Group 5 needs to first be verified and properly age-dated to establish a connection}
\vspace*{0.1in}
\end{deluxetable*}

%% file: AGES.tex
\begin{deluxetable*}{cccccc}
\tablecolumns{6}
\tablewidth{0pt}
\tabletypesize{\scriptsize}
\tablecaption{Summary of age solutions for the 4 well-established components of the Austral Complex. All ages are in Myr.}
\label{tab:ages}
\tablehead{
\colhead{Group} &
\colhead{Dynamical} &
\multicolumn{3}{c}{Isochronal} &
\colhead{Adopted} \\
\colhead{} &
\colhead{} &
\colhead{PARSEC} &
\colhead{BHAC15} &
\colhead{DSEP-Magnetic} &
\colhead{}
}
\startdata
COL (1) &     25.2 $\pm$ 5.4 &    38.5 $\pm$ 3.4 &    14.6 $\pm$ 0.8 &    26.7 $\pm$ 2.0 &         26.4 $\pm$1.7 \\
CAR (2) &     32.9 $\pm$ 8.2 &    37.0 $\pm$ 2.6 &    14.5 $\pm$ 0.7 &    27.7 $\pm$ 2.1 &         26.4 $\pm$ 1.7 \\
THA (3) &     27.2 $\pm$ 5.9 &    53.0 $\pm$ 1.7 &    20.5 $\pm$ 0.9 &    47.2 $\pm$ 2.4 &         46.3 $\pm$ 2.3 \\
FH (4) &     25.0 $\pm$ 2.0 &    46.0 $\pm$ 2.1 &    16.7 $\pm$ 0.7 &    31.1 $\pm$ 1.5 &         32.4 $\pm$ 1.3 \\
\enddata
\vspace*{0.1in}
\end{deluxetable*}

%% file: binaries.tex
\begin{deluxetable*}{ccccccc} %%%note: source file is AUS_BIN_CATALOG.tex
\tablecolumns{7}
\tablewidth{0pt}
\tabletypesize{\scriptsize}
\tablecaption{Binaries in the Austral complex, including both the new sample in FH and the literature candidates in other Austral sub-associations. Objects identified as members of the same system are given the same system ID. The complete version of this table is available in the online-only version, which contains 206 stars}
\label{tab:binaries}
\tablehead{
\colhead{Gaia ID} &
\colhead{Sys ID} &
\colhead{$R$\tablenotemark{a}} &
\colhead{RA} &
\colhead{Dec} &
\colhead{m$_G$} &
\colhead{$\pi$} \\
\colhead{} &
\colhead{} &
\colhead{AU} &
\colhead{(deg)} &
\colhead{(deg)} &
\colhead{} &
\colhead{(mas)}
}
\startdata
 5081415806929560704 &          0 &     0 &   55.0453 & -27.3686 &  16.64 &      7.08 \\
 5081415806929560448 &          0 &   644 &   55.0449 & -27.3674 &  16.74 &      7.07 \\
 4863772118776214400 &          1 &     0 &   56.1201 & -30.9173 &   9.10 &      7.12 \\
 4863772118776215040 &          1 &  3884 &   56.1146 & -30.9235 &  16.76 &      7.18 \\
 4855535127416838016 &          2 &     0 &   57.1707 & -39.4095 &  13.28 &      7.36 \\
 4855535127416837888 &          2 &   643 &   57.1699 & -39.4083 &  14.45 &      7.38 \\
 4870445020485171712 &          3 &     0 &   63.8881 & -34.1556 &   9.02 &      7.39 \\
 4870445016189720192 &          3 &   860 &   63.8903 & -34.1556 &  17.17 &      7.84 \\
 5072131427664000640 &          4 &   901 &   45.8619 & -26.6121 &  18.78 &      7.99 \\
 5072131427665158272 &          4 &     0 &   45.8620 & -26.6123 &  18.45 &      7.20 \\
 5082759650657410560 &          5 &     0 &   58.9332 & -25.4919 &  13.98 &      8.06 \\
 5082759650657372160 &          5 &  4542 &   58.9305 & -25.5018 &  17.66 &      8.07 \\
\enddata
\tablenotetext{a}{Separation at the distance of the primary relative to the primary. Primaries have a separation of zero.}
\vspace*{0.1in}
\end{deluxetable*}